\documentclass[10pt, conference, letterpaper]{IEEEtran}
\usepackage[utf8]{inputenc}
\usepackage[english]{babel}
\usepackage[ruled, vlined]{algorithm2e}
\usepackage{algpseudocode}
\usepackage{amssymb}
\usepackage{array}
\usepackage{arydshln}
\usepackage{blkarray, bigstrut}
\usepackage{bm}
\usepackage{booktabs}
\usepackage{braket}
\usepackage{cite}
\usepackage{cuted}
\usepackage{enumitem}
\usepackage{graphicx}
\usepackage{gensymb}
\usepackage[mathscr]{euscript}
\usepackage[switch]{lineno}
\usepackage{mathtools}
\usepackage{multirow}
\usepackage{multicol}
\usepackage{mathtools}
\usepackage{xfrac}
\usepackage{framed} 
\usepackage[framed]{ntheorem}
\usepackage{pgfplots}
\usepgfplotslibrary{polar}
\usepackage{qtree}
\usepackage[outline]{contour}
\usepackage{adjustbox}
\usepackage{rotating}
\pgfplotsset{compat = 1.12}
\usepackage{stfloats}
\usepackage{subcaption}
\usepackage{url}
\usepackage{tikz}
\usepackage{tkz-berge}
\usepackage{fixltx2e}
\usepackage[framemethod=TikZ]{mdframed}
\usetikzlibrary{patterns}
\usetikzlibrary{spy}
\usetikzlibrary{shapes, backgrounds,calc, snakes}
\usetikzlibrary{decorations.pathreplacing}
\usetikzlibrary{matrix}
\usepackage{fancybox}
\usepackage{soul, xcolor}
\DeclarePairedDelimiter{\ceil}{\lceil}{\rceil}
\def\approxprop{%
  \def\p{%
    \setbox0=\vbox{\hbox{$\propto$}}%
    \ht0=0.6ex \box0 }%
  \def\s{%
    \vbox{\hbox{$\sim$}}%
  }%
  \mathrel{\raisebox{0.7ex}{%
      \mbox{$\underset{\s}{\p}$}%
    }}%
}

\tikzstyle{vertex} = [circle, draw, inner sep = 0pt, minimum size = 10pt]

\newframedtheorem{frm-prop}{Property}

\definecolor{bblue}{rgb}{0.12392, 0.0490, 0.9588}
\definecolor{sskyblue}{rgb}{0.1529, 0.5882, 0.9216}
\definecolor{ggreen}{rgb}{0.7098, 0.95, 0.40781}
\definecolor{yyellow}{rgb}{0.9765, 0.9804, 0.0784}

\definecolor{color0}{HTML}{FF0147}
\definecolor{color1}{HTML}{F400DC}
\definecolor{color2}{HTML}{BA0DFF}
\definecolor{color3}{HTML}{5700E8}
\definecolor{color4}{HTML}{0B03FF}
\definecolor{color5}{HTML}{0957F4}
\definecolor{color6}{HTML}{03B3FF}
\definecolor{color7}{HTML}{08E8DA}
\definecolor{color8}{HTML}{07FF8E}
\definecolor{color9}{HTML}{51FF0A}

\definecolor{p1}{rgb}{1, 0.0667, 0}
\definecolor{p2}{rgb}{1, 0.24, 0}
\definecolor{p3}{rgb}{1, 0.349, 0}
\definecolor{p4}{rgb}{1, 0.490, 0}
\definecolor{p5}{rgb}{1, 0.631, 0}
\definecolor{p6}{rgb}{1, 0.792, 0}
\definecolor{p7}{rgb}{1, 0.933, 0}

\definecolor{ccolor0}{HTML}{FF007D}
\definecolor{ccolor1}{HTML}{760CE8}
\definecolor{ccolor2}{HTML}{0A55FF}
\definecolor{ccolor3}{HTML}{0DB6F4}
\definecolor{ccolor4}{HTML}{00FF76}
\definecolor{ccolor5}{HTML}{6FE80C}
\definecolor{ccolor6}{HTML}{FFDE0A}
\definecolor{ccolor7}{HTML}{FF990A}

\definecolor{bcolor0}{HTML}{fff7fb}
\definecolor{bcolor1}{HTML}{ece7f2}
\definecolor{bcolor2}{HTML}{d0d1e6}
\definecolor{bcolor3}{HTML}{a6bddb}
\definecolor{bcolor4}{HTML}{74a9cf}
\definecolor{bcolor5}{HTML}{3690c0}
\definecolor{bcolor6}{HTML}{0570b0}
\definecolor{bcolor7}{HTML}{045a8d}
\definecolor{bcolor8}{HTML}{023858}

\definecolor{gcolor0}{HTML}{cce7e3}
\definecolor{gcolor1}{HTML}{b3dad4}
\definecolor{gcolor2}{HTML}{99cec6}
\definecolor{gcolor3}{HTML}{80c2b8}
\definecolor{gcolor4}{HTML}{67b6aa}
\definecolor{gcolor5}{HTML}{4daa9c}
\definecolor{gcolor6}{HTML}{349d8d}
\definecolor{gcolor7}{HTML}{1a917f}
\definecolor{gcolor8}{HTML}{018571}
\definecolor{gcolor9}{HTML}{017866}
\definecolor{gcolor10}{HTML}{016a5a}
\definecolor{gcolor11}{HTML}{015d4f}
\definecolor{gcolor12}{HTML}{015044}
\definecolor{gcolor13}{HTML}{014339}
\definecolor{gcolor14}{HTML}{00352d}
\definecolor{gcolor15}{HTML}{002822}
\definecolor{gcolor16}{HTML}{001b17}

\definecolor{gcolor7}{HTML}{474747}

\definecolor{qscolor0}{HTML}{fbb4ae}
\definecolor{qscolor1}{HTML}{b3cde3}
\definecolor{qscolor2}{HTML}{ccebc5}
\definecolor{qscolor3}{HTML}{decbe4}
\definecolor{qscolor4}{HTML}{fed9a6}
\definecolor{qscolor5}{HTML}{ffffcc}
\definecolor{qscolor6}{HTML}{e5d8bd}
\definecolor{qscolor7}{HTML}{fddaec}
\definecolor{qscolor8}{HTML}{f2f2f2}

\definecolor{qmcolor0}{HTML}{8dd3c7}
\definecolor{qmcolor1}{HTML}{ffffb3}
\definecolor{qmcolor2}{HTML}{bebada}
\definecolor{qmcolor3}{HTML}{fb8072}
\definecolor{qmcolor4}{HTML}{80b1d3}
\definecolor{qmcolor5}{HTML}{fdb462}
\definecolor{qmcolor6}{HTML}{b3de69}
\definecolor{qmcolor7}{HTML}{fccde5}
\definecolor{qmcolor8}{HTML}{d9d9d9}

\definecolor{dcolor0}{HTML}{313695}
\definecolor{dcolor1}{HTML}{4575b4}
\definecolor{dcolor2}{HTML}{74add1}
\definecolor{dcolor3}{HTML}{abd9e9}
\definecolor{dcolor4}{HTML}{e0f3f8}

\definecolor{dcolor5}{HTML}{fef0d9}
\definecolor{dcolor6}{HTML}{fdd49e}
\definecolor{dcolor7}{HTML}{fdbb84}
\definecolor{dcolor8}{HTML}{fc8d59}
\definecolor{dcolor9}{HTML}{ef6548}
\definecolor{dcolor10}{HTML}{d7301f}
\definecolor{dcolor11}{HTML}{990000}

\usepackage{todonotes}

\newcommand{\fref}[1]{Fig.~\ref{#1}}

\newcommand{\tref}[1]{Table~\ref{#1}}

\DeclareMathSizes{10pt}{8pt}{6pt}{5pt} 	

\begin{document}
\title{\LARGE{HydraWave: Multi-Group Multicast Hybrid Precoding and Low-Latency Scheduling for Ubiquitous Industry 4.0 mmWave Communications}}


\author{
\IEEEauthorblockN{Luis F. Abanto-Leon, Matthias Hollick, and Gek Hong (Allyson) Sim} 
\IEEEauthorblockA{Secure Mobile Networking (SEEMOO) Lab, Technichal University of Darmstadt, Germany}
}

\thanks{M. Shell was with the Department
of Electrical and Computer Engineering, Georgia Institute of Technology, Atlanta,
GA, 30332 USA e-mail: (see http://www.michaelshell.org/contact.html).}
\thanks{J. Doe and J. Doe are with Anonymous University.}
\thanks{Manuscript received April 19, 2005; revised August 26, 2015.}

\markboth{Journal of \LaTeX\ Class Files,~Vol.~14, No.~8, August~2015}%
{Shell \MakeLowercase{\textit{et al.}}: Bare Demo of IEEEtran.cls for IEEE Communications Society Journals}

\setstcolor{red}
\maketitle

\begin{abstract}
Industry 4.0 anticipates massive interconnectivity of industrial devices (e.g., sensors, actuators) to support factory automation and production. Due to the rigidity of wired connections to harmonize with automation, wireless information transfer has attracted substantial attention. However, existing solutions for the manufacturing sector face critical issues in coping with the key performance demands: \emph{ultra-low latency, high throughput, and high reliability}. Besides, recent advancements in wireless millimeter-wave technology advocates hybrid precoding with affordable hardware and outstanding spatial multiplexing performance. Thus, we present \texttt{\small{HYDRAWAVE}} --- a new paradigm that contemplates the joint design of \emph{group scheduling} and \emph{hybrid precoding} for multi-group multicasting to support ubiquitous low-latency communications. Our hybrid precoder, based on semidefinite relaxation and Cholesky matrix factorization, facilitates the robust design of the constant-modulus phase shifts rendering formidable performance at a fraction of the power required by fully-digital precoders. Further, our novel group scheduling formulation minimizes the number of scheduling windows while accounting for the channel correlation of the co-scheduled multicast receivers. Compared to exhaustive search, which renders the optimal scheduling at high overhead, \texttt{\small{HYDRAWAVE}} incurs only $ 9.5 \% $ more delay. Notoriously, \texttt{\small{HYDRAWAVE}} attains up to $ 102 \% $ gain when compared to the other benchmarked schemes.
\end{abstract}


\IEEEpeerreviewmaketitle

\section{Introduction}

The industrial revolution, Industry 4.0, fosters smart factories of the future where the components in a production chain---such as industrial equipment, logistics, products, and processes---are inherently interconnected. Given the hyper-connected vision of Industry 4.0, wired connections become less attractive because they (\emph{i}) hinder automation by constraining the movement of industrial robotics and (\emph{ii}) slow down mechanics. To overcome this obstacle, in recent years, significant effort has been dedicated to leveraging the benefits of wireless communications solutions for the manufacturing sector (e.g., 6TiSCH, ZigBee). However, the adoption of these technologies is either limited or has not been consolidated due to the uncertainty on their capability to offer performance similar to optical fiber and guarantee the critical requirements of industrial applications: \emph{ultra-high throughput and ultra-low latency with emphasis on high reliability} \cite{b6}. 

Owing to recent advancements in wireless millimeter-wave (mmWave) technology, the mmWave spectrum is regarded as a solution to wire replacement in industrial/manufacturing sectors. In addition to providing multi-Gbps rates due to wideband availability, mmWave frequencies characterize for requiring antennas with a small footprint that can be easily embedded onto miniature machinery/devices. A measurement campaign conducted in an industrial environment has concluded that mmWave communications is feasible and effortlessly implementable in such environments \cite{b7}. Nevertheless, the research on wireless and (in particular) mmWave technology for the industrial context is still at its infancy. 
\begin{figure*}[t!]
\centering
\begin{subfigure}{0.38\textwidth}
	\centering
    \includegraphics[width=0.75\columnwidth]{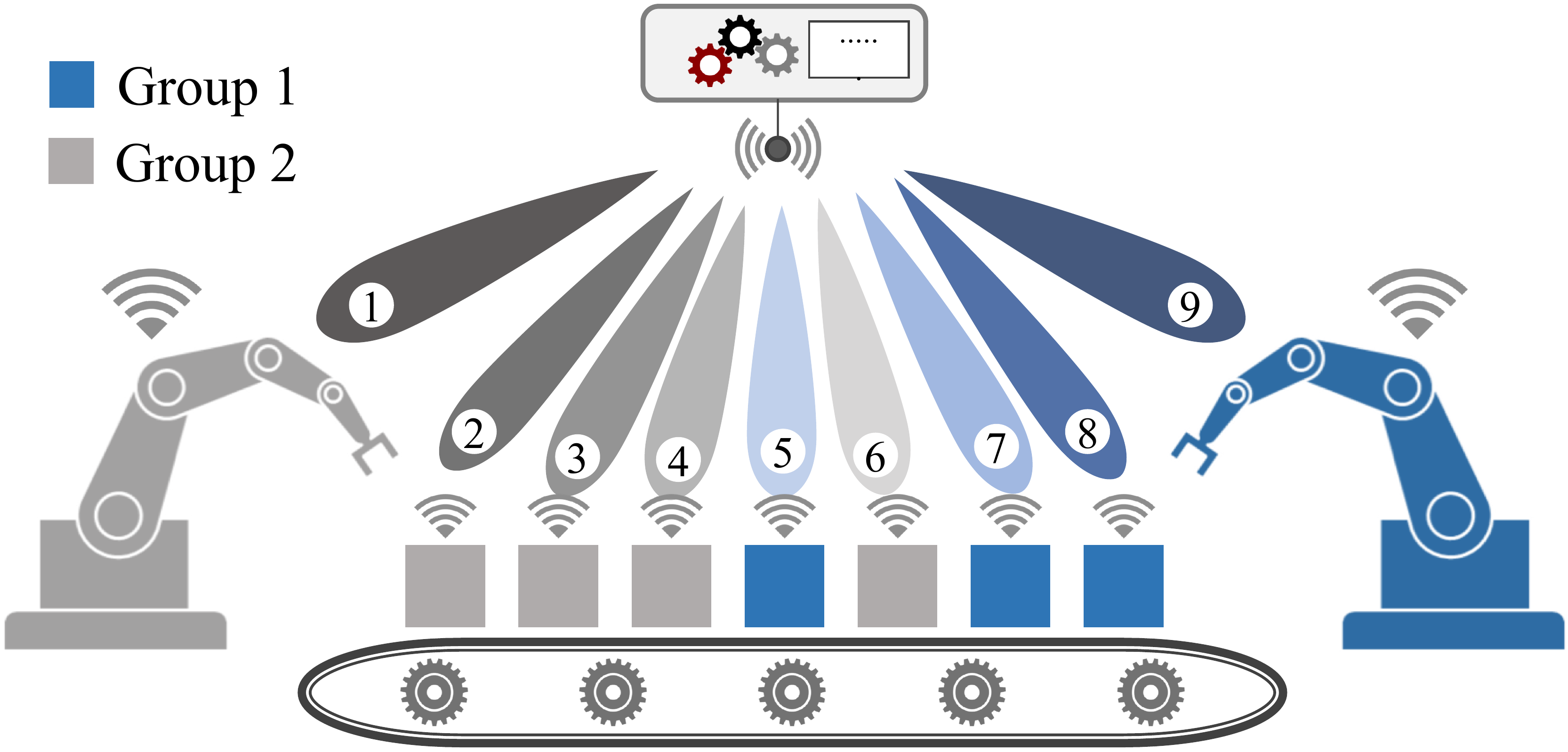}
    \vspace{-2mm}
	\caption{Conventional method with sequential unicast}
	\label{fig:singlegroup}
\end{subfigure}
\hspace{2mm}
\begin{subfigure}{0.38\textwidth}
	\centering
    \includegraphics[width=0.75\columnwidth]{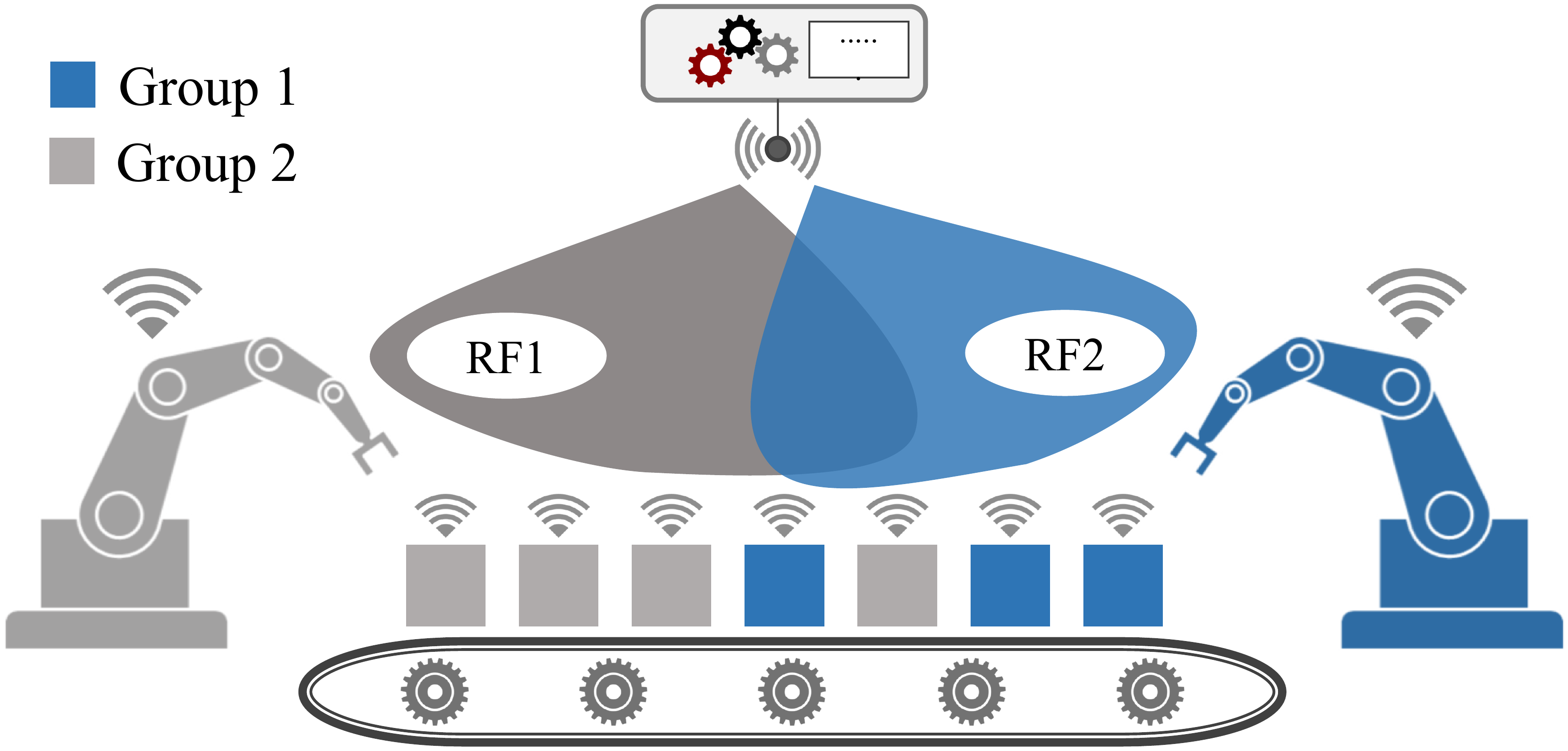}
    \vspace{-2mm}
	\caption{Multi-group multicasting}
	\label{fig:multigroup}
\end{subfigure}
\hspace{2mm}
\begin{subfigure}{0.18\textwidth}
	\centering
    \includegraphics[width=0.6\columnwidth]{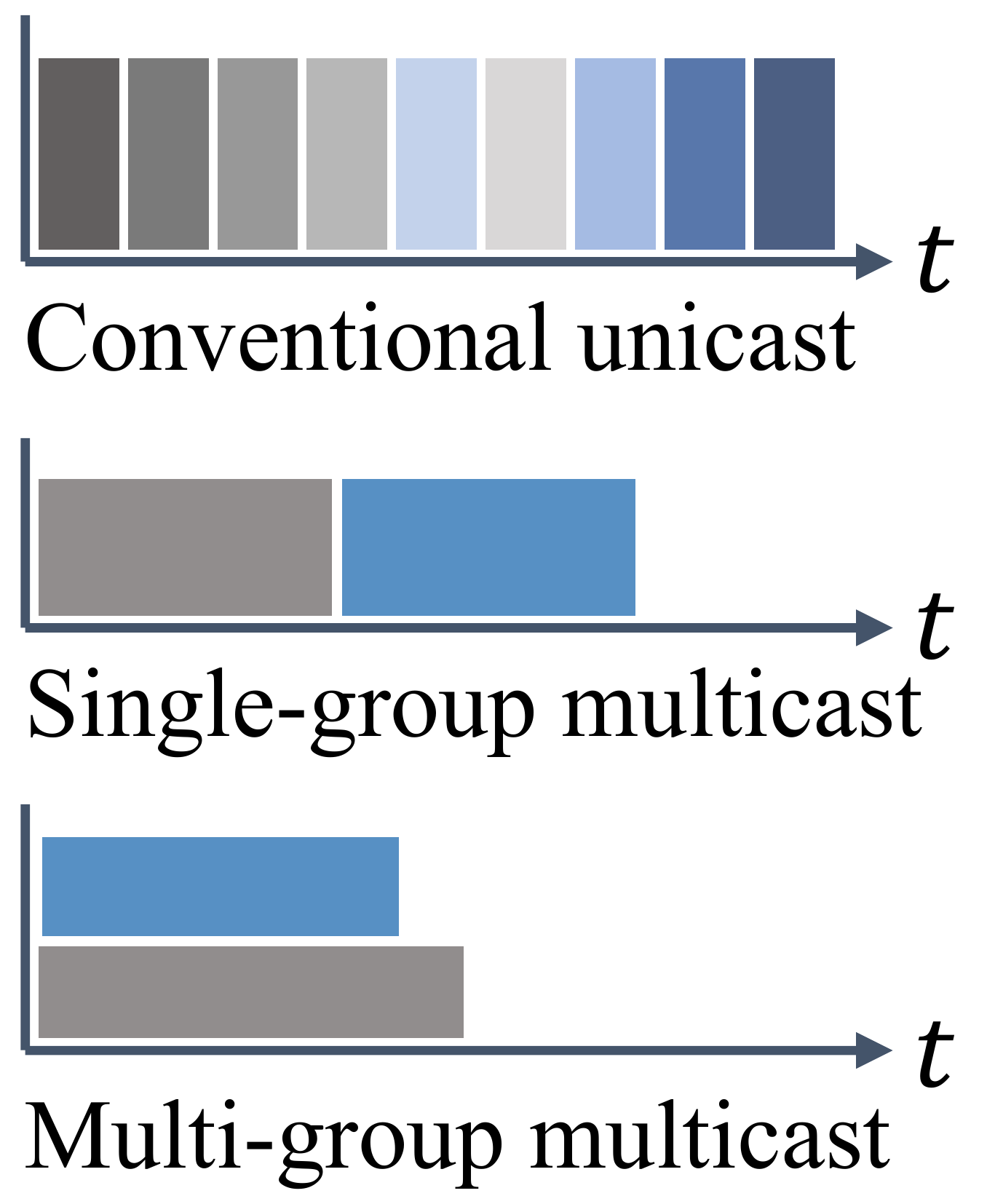}
    \vspace{-2mm}
    \caption{Latency performance}
    \label{fig:latency}
\end{subfigure}
\vspace{-2mm}
\caption{Comparison of mmWave unicast, conventional and proposed multi-group multicasting methods in Industry 4.0.}
\label{fig:multicast}
\vspace{-7mm}
\end{figure*}

The industrial sectors can profoundly benefit from mmWave multicast beamforming to deliver common information to industrial equipment (e.g., sensors and actuators) for the distributed organization of production. In such setting, instead of sequentially transmitting unicast streams from a transmitter to each receiver (\fref{fig:singlegroup}), multicasting to a group of devices (\fref{fig:multigroup}) can boost the spectral efficiency and offer lower latency. Moreover, if the transmitter is equipped with multiple radio frequency (RF) chains, multi-group multicasting can be enabled thus making possible to serve several groups of devices concurrently while further leveraging the gains in terms of spectral efficiency and latency (\fref{fig:latency}). 

We envision a heterogeneous hyper-connected Industry 4.0 with \emph{(i) anchored receivers} (e.g., sensors, actuators, programmable logic devices) that perform tasks locally and \emph{(ii) mobile receivers} (e.g., robots) that carry out tasks at various locations. Specifically, multiple \emph{anchored receivers} from different multicast groups can be found sharing the same space, thus increasing the difficulty of spatial multiplexing (\fref{fig:multigroup}). In addition, due to mobility (e.g. robots), numerous \emph{ mobile receivers} with different information needs may temporarily change their location to carry out specific tasks, thus either altering the density of devices at different phases of the production chain or generating a variable degree of interference.


Fully-digital precoders with a massive amount of antennas and RF chains are a tangible technology at sub-6GHz frequencies. However, in mmWave, these precoders are still not affordable due to hardware complexity and high power consumption. As a result, substantial effort has been oriented to \emph{(i)} improve the architectures design (e.g., \cite{b8, b9}) and \emph{(ii)} develop accurate channel estimation methods (e.g., \cite{b10, b11}) for hybrid precoders, aiming at facilitating physical-layer hybrid precoding with high signal-to-interference-plus-noise ratio (SINR). In industrial scenarios, the number of multicast groups is expected to be large in comparison to the number of RF chains, generating thereby the necessity for scheduling. Further, depending on how the groups are co-scheduled, the achievable performance of the precoder will be impacted. As a result, a scrupulous design that considers the joint optimization of scheduling, hybrid precoder (at the transmitter) and analog combiners (at receivers) is required to comply with the requirements of industrial applications. To this end, we propose \texttt{HYDRAWAVE}, a versatile scheme that devises an \emph{agile scheduler} and a \emph{robust multi-group multicast precoder with analog combiners}, capable of providing \emph{high throughput, low latency, and high reliability} (which is guaranteed through service ubiquitousness).


\begin{table}[!h]
	\fontsize{6}{7}\selectfont
	\centering
	\setlength\tabcolsep{2pt}
	\renewcommand{\arraystretch}{0.25}
	\centering
	\caption{Literature on multicasting}
	\label{t1}
	\begin{tabular}{m{0.4cm} m{1.8cm} @{}c @{}c @{}c @{}c}
		\toprule
		\multirow{3}{0.4cm}{\centering{Type}} &
		\multirow{3}{1.8cm}{\centering{Scheduling with PHY abstraction}} & 
		\multicolumn{2}{c}{PHY design w/o scheduling} & 
		\multicolumn{2}{c}{PHY design w/ scheduling} \\ 
		\cmidrule(r){3-6}
		& & Fully-digital & Hybrid & Fully-digital & Hybrid \\ 
		\midrule
		\midrule
		\centering{SGM} & 
		\parbox{1.4cm}{\centering {\cite{b38, b39, b40}}} & 
		\parbox{1.2cm}{\centering {\cite{b30, b31, b32, b33, b34}}} & 
		\parbox{1.4cm}{\centering {\cite{b32, b35, abanto2020:learning-maxmin-hybrid-precoding-mmWave-multicasting}}} & 
		\parbox{1.2cm}{\centering {\cite{b36, b37}}} & 
		\parbox{1.2cm}{\centering {--}} \\
		
		\midrule
		
		\centering{MGM} & 
		\parbox{1.4cm}{\centering {--}} & 
		\parbox{1.2cm}{\centering {\cite{b19, b20, b21, b24, b25, b27, b28, b29}}} & 
		\parbox{1.2cm}{\centering {\cite{b22, b23, b26, abanto2019:multigroup-multicast-hybrid-precoding}}} & 
		\parbox{1.2cm}{\centering {--}} & 
		\parbox{1.2cm}{\centering {--}} \\
		
		\bottomrule
	\end{tabular}
	\vspace{-2mm}
\end{table}

As shown in Table \ref{t1}, the literature on single-group multicasting (SGM) focus on either pure scheduling with physical-layer (PHY) abstraction \cite{b38, b39, b40} or physical-layer beamforming/precoding with \cite{b36, b37} and without scheduling \cite{b30, b31, b32, b33, b34, b35, abanto2020:learning-maxmin-hybrid-precoding-mmWave-multicasting}, ignoring the interference aspect in multi-group multicast scenarios. On the other hand, the body of research on multi-group multicasting (MGM) focuses mostly on fully-digital precoder designs without scheduling tailored for sub-6GHz frequencies \cite{b19, b20, b21, b24, b25, b27, b28, b29}, which are not appropriate for mmWave. As a result, the more practical and cost-efficient hybrid precoders \cite{b22, b23, b26} have attracted substantial attention. Nevertheless, these solutions are either (\emph{i}) constrained in application due to simplified assumptions \cite{b21, b23} or (\emph{ii}) unimplementable in the existing multi-antenna hardware due to customized designs \cite{b22, b26}. In \cite{abanto2019:multigroup-multicast-hybrid-precoding}, MGM is investigated with hybrid precoders and finite resolution phase shifts but the scheduling aspect is not considered. The joint problem of \emph{group scheduling} and \emph{precoding} for multi-group multicasting has been studied for neither fully-digital nor hybrid precoders. The following summarizes our contributions: 
\begin{itemize}
	\item We introduce three propositions that support the design of precoders that ensure low-latency and high reliability. \emph{Proposition 1} renders insights on the (approximate) inverse relation between SINR and latency. \emph{Proposition 2} leverages on this result, and reveals that latency minimization is promoted when maximizing the minimum equalized SINR (e-SNR). This exposes a relevant relation with the max-min MGM problem. \emph{Proposition 3} supports our formulation of optimal joint group scheduling and precoding for the multi-group multicast problem, with emphasis on latency minimization and high reliability.
	
	\item Due to the complexity of solving the problem related to \emph{Proposition 3} (which requires exhaustive search), we devise a novel group scheduling formulation with a two-fold objective: \emph{(i)} minimizing the number of scheduling windows and \emph{(ii)} reducing channel correlation between the co-scheduled receivers. To attain the latter objective, we introduce a metric called \emph{aggregate inter-group correlation} that is shown to be highly suitable for forming the co-scheduled multicast groups. As a result, we only need to solve the problem associated with \emph{Proposition 2} (which is comparatively simpler) for every window of the resultant scheduling thus reducing the complexity.
	
	\item The problem associated with \emph{Proposition 2} is non-convex. Due to the NP-hardness of this problem, we propose an alternate optimization scheme where the analog and digital components of the hybrid precoder (at the transmitter), and the analog combiners (at each receiver) are optimized sequentially. We recast each sub-problem as a semidefinite relaxation (SDR) program in order to convexify the non-convex expressions.
	
	\item We propose a versatile approach based on Cholesky matrix factorization, capable of handling an arbitrary number of phase rotations at both the analog component of the hybrid precoder (at the transmitter) and the analog combiners (at the receivers) with outstanding performance.
	
	\item Compared to prior art on multi-group multicast precoding, we consider receivers with multiple antennas, which can be adopted at mmWave frequencies owing to the small antenna footprints.
	
	\item Through extensive simulations, we evaluate the performance of our hybrid precoder and compare it against fully-digital and fully-analog implementations. Further, we also assess our proposed \texttt{\small{HYDRAWAVE}} (joint group scheduling and precoding) and compare it in terms of \emph{latency} against three competing approaches: single-group multicasting (\texttt{\small{SING}}), random scheduling (\texttt{\small{RAND}}) and exhaustive search (\texttt{\small{XHAUS}}). We show through simulations that \texttt{\small{HYDRAWAVE}} can attain gains up to $ 102 \% $ and $ 60 \% $ when compared against \texttt{\small{RAND}} and \texttt{\small{SING}}, respectively, while remaining within $ 9.5 \% $ optimality of \texttt{\small{XHAUS}}.
	
\end{itemize}

\section{Related Work}
\label{sec:related_work}
In SGM scheduling with physical-layer abstraction \cite{b38, b39, b40}, the transmitter adjusts the gains of single-lobe switched beams to improve the SNR at the receivers. These works mainly focus on achieving high throughput with minimum delay for all the multicast receivers. Concerning SGM physical-layer precoding (hybrid/fully-digital and with/without scheduling), researchers have studied the quality-of-service (QoS) and max-min fairness problems \cite{b30, b31, b32, b33, b34, b35, b36, b37}. Under these two categories, the QoS \cite{b19, b20, b21, b22, b23, b24, b25} and max-min fairness \cite{b24, b25, b26, b27, b28, b29} problems have also been researched for the MGM case. Fully-digital precoders are highly versatile for interference mitigation due to the availability of numerous RF chains. However, these designs consumes excessive power and require expensive hardware (particularly for mmWave). To address these issues, the use of analog-digital architectures (i.e., hybrid precoders) have received considerable attention. Hybrid precoders are constituted by a low-dimensional digital precoder that allows interference management and a high-dimensional network of phase shifters, with an arbitrary set of phase rotations that facilitates beamsteering. These designs do not have the same versatility as fully-digital implementations but are more energy-efficient and practical. Nevertheless, the existing hybrid precoder solutions (in general, multicast and multi-user unicast designs) are restricted in usage due to either (\emph{i}) simplified assumptions in phase rotations selection or (\emph{i}) unimplementability owing to customized hardware. Specifically, the solution propounded in \cite{b26} requires a especially connected network of phase shifters for optimal operation. On the other hand, in \cite{b21, b23}, the usage is constrained to only four different phase shifts. In \cite{b22}, the analog phase shifters are replaced by high-resolution lens arrays with adjustable power, thus circumventing the actual problem of phase selection.

A work that focuses on a problem slightly related to ours is \cite{b42}, where the authors study user selection and MGM precoding with fully-digital transmitters and single-antenna receivers. In \cite{b42}, specific users for each multicast group are selected in order to maximize the sum-rate. Contrastingly, in our case we deal with group selection for which we devise a metric (i.e., IGC) that depicts the mean channel vector correlation among the receivers of each group. Also, in our case, the objective is to design the precoders aiming at the minimization of the transmission latency.

\section{Multi-Group Multicast System Model}
\label{sec:model}
We consider a mmWave system where a next generation Node B (gNodeB) equipped with a hybrid precoder aims to schedule $ G_T $ multicast groups comprising a total of $ K_T $ receivers. The sets of receivers and groups are denoted by $ \mathcal{K} = \left\lbrace 1, 2, \dots, K_T \right\rbrace $ and $ \mathcal{I} = \left\lbrace 1, 2, \dots, G_T \right\rbrace $, respectively. In addition, $ \mathcal{G}_i $ represents the set of receivers in multicast group $ i \in \mathcal{I} $. The number of receivers in the $ i $-th multicast group is represented by $ \left| \mathcal{G}_i \right| $. As in \cite{b24}, we assume that $ \mathcal{G}_i \cap \mathcal{G}_{i'} = \left\lbrace \emptyset \right\rbrace, \forall i \neq i' $. The hybrid precoder at the gNodeB exhibits a sub-connected architecture, which consists of $ N_\mathrm{tx} $ transmit antennas and $ N^\mathrm{RF}_\mathrm{tx} $ RF chains. Thus, each RF chain is connected to a sub-array of $ L_\mathrm{tx} = N_\mathrm{tx} / N^\mathrm{RF}_\mathrm{tx} $ antennas \cite{b8}. Due to capacity constraints, each receiver is equipped with purely analog combiners that consist of $ N_\mathrm{rx} $ antennas and a single RF chain, i.e, $ N^\mathrm{RF}_\mathrm{rx} = 1 $ and $ L_\mathrm{rx} = N_\mathrm{rx} $. Since, in general $ G_T \geq N^\mathrm{RF}_\mathrm{tx} $, scheduling is necessary because $ N^\mathrm{RF}_\mathrm{tx} $ determines the maximum number of data streams (or groups) that can be spatially multiplexed.

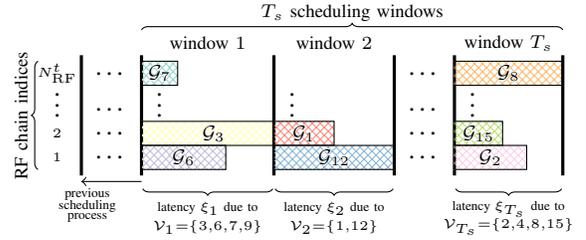
\begin{figure}[!t]
	\centering
	\begin{tikzpicture}[scale = 0.8]
	
	\draw[decoration = {brace, raise=5pt}, decorate] (-1.1,-0.4) -- node[right = 6pt] {} (-1.1,1.4);
	\node[rotate = 90] at (-1.6,0.5) {\scriptsize RF chain indices};
	
	\node at (-1.0,1.2) {\tiny $ N^t_{\mathrm{RF}} $};
	\node at (-1.0,0.8) {\vdots};
	\node at (-1.0,0.2) {\tiny $ 2 $};
	\node at (-1.0,-0.2) {\tiny $ 1 $};
	
	\draw[very thick] (-0.6,1.5) -- (-0.6,-0.5);
	
	\node at (-0.1,1.2) {\dots};
	\node at (-0.1,0.6) {\dots};
	\node at (-0.1,0.2) {\dots};
	\node at (-0.1,-0.2) {\dots};
	
	\draw[very thick] (0.4,1.5) -- (0.4,-0.5);

	\draw[pattern = crosshatch, pattern color = qmcolor0!90] (0.4,1.4) rectangle (1.0,1.0);
	\node at (0.7,0.8) {\vdots};
	\draw[pattern = crosshatch, pattern color = qmcolor1!90] (0.4,0.4) rectangle (2.6,0.0);
	\draw[pattern = crosshatch, pattern color = qmcolor2!90] (0.4,0) rectangle (1.8,-0.4);
	
	\node at (0.7,1.2) {\contour{white}{\scriptsize $ \mathcal{G}_7 $}};
	\node at (1.6,0.2) {\contour{white}{\scriptsize $ \mathcal{G}_3 $}};
	\node at (1.1,-0.2) {\contour{white}{\scriptsize $ \mathcal{G}_6 $}};
	
	\draw[decoration = {brace, raise=5pt}, decorate] (2.58,-0.5) -- node[right = 6pt] {} (0.42,-0.5);
	\node at (1.5,-1.05) {\tiny latency $ \xi_1 $ due to};
	\node at (1.5,-1.35) {$ \scriptscriptstyle \mathcal{V}_1 = \left\lbrace 3, 6, 7, 9 \right\rbrace $}; 
	
	\draw[very thick] (2.6,1.5) -- (2.6,-0.5);
	
	\draw[pattern = crosshatch, pattern color = qmcolor3!90] (2.6,0.4) rectangle (3.6,0.0);
	\draw[pattern = crosshatch, pattern color = qmcolor4!90] (2.6,0) rectangle (4.6,-0.4);
	
	\node at (2.9,0.8) {\vdots};
	\node at (3.1,0.2) {\contour{white}{\scriptsize $ \mathcal{G}_1 $}};
	\node at (3.6,-0.2) {\contour{white}{\scriptsize $ \mathcal{G}_{12} $}};
	
	\draw[decoration = {brace, raise=5pt}, decorate] (4.58,-0.5) -- node[right = 6pt] {} (2.62,-0.5);
	\node at (3.6,-1.05) {\tiny latency $ \xi_2 $ due to};
	\node at (3.6,-1.35) {$ \scriptscriptstyle \mathcal{V}_2 = \left\lbrace 1, {12} \right\rbrace $}; 

	\draw[very thick] (4.6,1.5) -- (4.6,-0.5);
	
	\node at (5.1,1.2) {\dots};
	\node at (5.1,0.6) {\dots};
	\node at (5.1,0.2) {\dots};
	\node at (5.1,-0.2) {\dots};
	
	\draw[very thick] (5.6,1.5) -- (5.6,-0.5);
	
	\draw[pattern = crosshatch, pattern color = qmcolor5!90] (5.6,1.4) rectangle (7.4,1.0);
	\node at (5.9,0.8) {\vdots};
	\draw[pattern = crosshatch, pattern color = qmcolor6!90] (5.6,0.4) rectangle (6.4,0.0);
	\draw[pattern = crosshatch, pattern color = qmcolor7!90] (5.6,0) rectangle (6.8,-0.4);
	
	\node at (6.5,1.2) {\contour{white}{\scriptsize $ \mathcal{G}_8 $}};
	\node at (6.0,0.2) {\contour{white}{\scriptsize $ \mathcal{G}_{15} $}};
	\node at (6.2,-0.2) {\contour{white}{\scriptsize $ \mathcal{G}_2 $}};
	
	\draw[very thick] (7.4,1.5) -- (7.4,-0.5);

	\draw[decoration = {brace, raise=5pt}, decorate] (7.38,-0.5) -- node[right = 6pt] {} (5.62,-0.5);
	\node at (6.5,-1.05) {\tiny latency $ \xi_{T_s} $ due to};
	\node at (6.5,-1.35) {$ \scriptscriptstyle \mathcal{V}_{T_s} = \left\lbrace 2, 4, 8, {15} \right\rbrace $}; 
	
	\node at (1.5,1.7) {\scriptsize window $ 1 $};
	\node at (3.6,1.7) {\scriptsize window $ 2 $};
	\node at (6.5,1.7) {\scriptsize window $ T_s $};
	
	\draw[decoration = {brace, raise=5pt}, decorate] (0.4,1.7) -- node[right = 6pt] {} (7.4,1.7);
	\node at (3.9,2.2) {\scriptsize $ T_s $ scheduling windows};
	
	\draw[->] (0.4,-0.6)--(-0.6,-0.6);
	\node at (-0.5,-0.8) {\tiny previous};
	\node at (-0.5,-1.0) {\tiny scheduling};
	\node at (-0.5,-1.2) {\tiny process};
	
	\end{tikzpicture}
	\vspace{-3mm}	
	\caption{Multicast groups scheduling: Multicast groups $ 1 $ and $ 12 $ have been jointly scheduled during window $ t = 2 $.}
	\label{f2}
	\vspace{-6mm}
\end{figure}

We define $ T_s $ as the number of scheduling windows, wherein mutually exclusive subsets of multicast groups are served, as shown in Fig \ref{f2}. The admissible range of scheduling windows, $ T_s $, depends on both $ N^\mathrm{RF}_\mathrm{tx} $ and $ G_T $. Specifically, $ \ceil[\big]{ \frac{G_T}{ N^\mathrm{RF}_\mathrm{tx} } } \leq T_s \leq G_T $ is maximum when exactly one multicast group is served per scheduling window. Also, $ T_s $ is minimum when the gNodeB operates at maximal utilization, serving $ N^\mathrm{RF}_\mathrm{tx} $ multicast groups simultaneously during every window (except for at most one window, if $ G_T \mod N^\mathrm{RF}_\mathrm{tx} \neq 0 $). Let $ \mathcal{V}_t $ denote an ordered set containing the indices of the co-scheduled groups within window $ t $. In addition, $ \mathcal{U}_t $ denotes the set of receivers catered during window $ t $, i.e. $ \mathcal{U}_t = \left\lbrace k \mid k \in \mathcal{G}_i, i \in \mathcal{V}_t \right\rbrace $. Similarly, $ \left\lbrace \mathcal{V}_t \right\rbrace^{T_s}_{t=1} $ and $ \left\lbrace \mathcal{U}_t \right\rbrace^{T_s}_{t=1} $ denote the collection of $ \mathcal{V}_t $ and $ \mathcal{U}_t $ for all scheduling windows, respectively. Since every multicast group is served only once within $ T_s $ windows, then $ \mathcal{V}_t \cap \mathcal{V}_{t' \neq t} = \left\lbrace \emptyset \right\rbrace $, $ \mathcal{U}_t \cap \mathcal{U}_{t' \neq t} = \left\lbrace \emptyset \right\rbrace $. Due to the multicast nature of the system, every receiver $ k \in \mathcal{G}_i $ requires the same length-$B_i$ bit stream $ \mathbf{b}_i = [ b_{i_1}, \dots b_{i_{B_i}} ]^T $. Thus, the amount of bits required by receiver $ k $ is $ B_k = B_i, \forall k \in \mathcal{G}_i $. At the transmitter, each bit stream $ \mathbf{b}_i $ is encoded at a suitable rate $ \pi_i $ that allows successful decoding at every intended receiver. As a result, $ G_T $ symbol streams $ \tilde{\mathbf{s}}_i = [ \tilde{s}_{i_1}, \tilde{s}_{i_2}, \dots ]^T $ ($ i \in \mathcal{I} $) with symbol-wise average unit power are produced, i.e., $ \mathbb{E} \left\lbrace \tilde{\mathbf{s}}_i \tilde{\mathbf{s}}^H_i \right\rbrace = \mathbf{I} $. The symbol streams are arranged in a matrix $ \mathbf{S} = \left[ \tilde{\mathbf{s}}_1, \dots, \tilde{\mathbf{s}}_{G_T} \right]^T  $ that is zero-padded where necessary, as the length of each stream $ \tilde{\mathbf{s}}_i $ depends on $ B_i $ and $ \pi_i $. Specifically, the non-zero entries of $ \mathbf{S} $ represent the effective data that needs to be delivered to all the groups (within $ T_s $ scheduling windows). Let $ \hat{\mathbf{S}}_t = \mathbf{V}_t \mathbf{S} $ denote the $ |\mathcal{V}_t| $ symbol streams transmitted during the $ t $-th scheduling window, where $ \mathbf{V}_t \in  \left\lbrace 0, 1 \right\rbrace^{| \mathcal{V}_t | \times G_{T}} $ is a binary scheduling matrix that filters the symbol streams from $ \mathbf{S} $. The \emph{transmission latency} associated to the delivery of $ \hat{\mathbf{S}}_t $ is defined as $ \xi_t = \max_{i \in \mathcal{V}_t } \frac{B_i}{\pi_i} $, which represent the minimal time interval required by the devices in all the multicast groups $ i \in \mathcal{V}_t $ to receive the intended data.

For any window $ t $, the analog and digital precoders of the hybrid transmitter are denoted by $ \mathbf{F}_t \in \mathbb{C}^{N_\mathrm{tx} \times N^\mathrm{RF}_\mathrm{tx}} $ and $ \mathbf{M}_t \in \mathbb{C}^{N^\mathrm{RF}_\mathrm{tx} \times | \mathcal{V}_t |} $, respectively. Any element $ \left( q,r \right)  $ of the analog precoder is a phase rotation with constant modulus. Thus, $ \left[ \mathbf{F}_t \right]_{q,r} \in \mathcal{F}$, where $ q \in \mathcal{Q} = \left\lbrace (r-1) L_\mathrm{tx} + l \mid 1 \leq l \leq L_\mathrm{tx} \right\rbrace $, $ r \in \mathcal{R} = \left\lbrace 1, \dots, N^\mathrm{RF}_\mathrm{tx} \right\rbrace $, and $~\mathcal{F} = \bigg\{ \sqrt{\delta_F} \dots, \sqrt{\delta_F} e^{j \frac{ 2 \pi \left( D_F - 1 \right) }{D_F}} \bigg\} $. $ D_F $ denotes the number of different phase shifts allowed at the transmitter, and $ \delta_F $ is a scaling factor. The combiner of the $ k $-th receiver is denoted by $ \mathbf{w}_k \in {\mathbb{C}}^{N_\mathrm{rx} \times 1}$, where $ \left[ \mathbf{w}_k \right]_{l} \in \mathcal{W}$, $ l \in \mathcal{L} = \left\lbrace 1, \dots, N_\mathrm{rx} \right\rbrace $, $ \mathcal{W} = \bigg\{ \sqrt{\delta_W}, \dots, \sqrt{\delta_W} e^{j \frac{ 2 \pi \left( D_W - 1 \right) }{D_W}} \bigg\} $, $ D_W $ is the number of phase shifts allowed at the analog combiner, and $ \delta_W $ is a scaling factor. The instantaneous downlink signal is represented by $ \mathbf{x}_t = \mathbf{F}_t \mathbf{M}_t \tilde{\mathbf{s}} $, where $ \tilde{\mathbf{s}} = \left[ \tilde{s}_{1_c}, \dots, \tilde{s}_{{| \mathcal{V}_t |}_c} \right]^T \in \mathbb{C}^{| \mathcal{V}_t | \times 1} $ sweeps through every column $ c $ of $ \hat{\mathbf{S}}_t $ extracting the multicast symbols to be delivered during window $ t $. The received signal at the $ k $-th device is denoted by $ y_k = \mathbf{w}^H_k \mathbf{H}_k \mathbf{x}_t $, where $ \mathbf{H}_k \in {\mathbb{C}}^{ N_\mathrm{rx} \times N_\mathrm{tx}} $ denotes the channel between the device and the gNodeB, whereas $ \mathbf{n}_k \sim \mathcal{CN} \left( \mathbf{0}, {\sigma}^2 \mathbf{I} \right) $ denotes additive white Gaussian noise. Assuming a flat-fading channel, $ y_k $ is given by
\begin{align} \label{e1}
	y_k = \underbrace{
					  	\mathbf{w}^H_k \mathbf{H}_k \mathbf{F}_t \mathbf{M}_t \mathbf{s}_{i_t}
		 			 }
		 			 _{\text{desired multicast signal}} +
		  \underbrace{
						\mathbf{w}^H_k \mathbf{H}_k \sum^{|\mathcal{V}_t|}_{\substack{j_t = 1, j_t \neq i_t}} \mathbf{F}_t \mathbf{M}_t \mathbf{s}_{j_t}
		 			 }
		 			 _{\text{interference}} + 
	 	  \underbrace{
						\mathbf{w}^H_k \mathbf{n}_k
		 			 }
		 			 _{\text{noise}}, 
\end{align}
where $ i_t \in \left\lbrace 1, 2, \dots, |\mathcal{V}_t| \right\rbrace $ is a relative index to denote the elements of $ \mathcal{V}_t $. Also, $ \mathbf{s}_{i_t} = \tilde{\mathbf{s}} - \mathbf{s}_{j_t} $ is a vector that contains zeros except for the $ i_t $-th position, which stores the $ i_t $-th element of $ \tilde{\mathbf{s}} $. Thus, the SINR at receiver $ k $ is defined as
\begin{equation} \label{e2}
	\mathrm{SINR}_k = 
			\frac
			{\left| \mathbf{w}^H_k \mathbf{H}_k \mathbf{F}_t \mathbf{M}_t \mathbf{e}_{i_t} \right|^2}
	 		{\displaystyle \sum^{|\mathcal{V}_t|}_{\substack{j_t = 1,  j_t \neq i_t }} \left| \mathbf{w}^H_k \mathbf{H}_k \mathbf{F}_t \mathbf{M}_t \mathbf{e}_{j_t} \right|^2 + \sigma^2 \left\| \mathbf{w}_k \right\|^2_2},  
\end{equation}
where $ \mathbf{e}_{i_t} $ stores a $ 1 $ in the $ i_t $-th position if $ k \in \mathcal{V}_t \left\lbrace i_t \right\rbrace $ or $ 0 $ otherwise. Based on \cite{b12}, the achievable rate $ \pi $ is approximated by the modified Shannon capacity,
\begin{equation} \label{e3}
	\pi = \min \{ C, C_{\mathrm{max}} \},
\end{equation}
where $ C = \log_2 (1 + \beta \cdot \mathrm{SINR}) $, $ C_{\mathrm{max}} $ is the capacity limit and $ \beta = 0.5 $ represents the SINR loss \cite{b12}. In (\ref{e1}) and (\ref{e2}), we have dropped the subscript $ t $ when referring to $ \mathbf{w}_k $ because only the combiners of the scheduled receivers need to be designed.

\noindent{\textbf{Remark:}} $ \mathbf{V}_t $ is obtained from $ \mathcal{V}_t = \left\lbrace v^t_1, v^t_2, \dots \right\rbrace $. For instance, if $ v^t_{a} \left( a \in \left\lbrace 1, 2, \dots, {|\mathcal{V}_t|} \right\rbrace \right) $ is in the $ z $-th position in $ \mathcal{V}_t $, then the element in the $ z $-th row and $ v^t_a $-th column of $ \mathbf{V}_t $ is set to $ 1 $.

\section{Scheduling Criteria for Multicasting }
Note that $ T_s $ is not necessarily related to the overall transmission latency $ \xi = \sum^{T_s}_{t = 1} \xi_t $. Each bit stream $ \mathbf{b}_i $ is to be transmitted at an optimal rate $ \pi_i $ that not only should provide service ubiquitousness (data decoding at every intended receiver) but also should minimize the transmission latency of the co-scheduled groups. Encoding every $ \mathbf{b}_i \in \mathcal{V}_t $ at a high rate will promote low latency $ \xi_t $, but it will require superior SINR at every receiver $ k \in \mathcal{G}_i ( i \in \mathcal{V}_t ) $ to ensure successful decoding. Nevertheless, under constrained transmit and receive power, a high SINR at every receiver cannot be guaranteed, compromising thereby the service ubiquitousness. On the other hand, transmitting at very low rate improves data decodability for a larger number of receivers (ubiquitousness enhancement) at the expense of increasing latency. In light of this observation, we realize that the optimal rate needs to be carefully devised to \emph{(i)} improve the ubiquitousness and \emph{(ii)} minimize the latency. However, the optimal rate depends on the SINR, which can only be determined once the scheduling, precoder and combiners have been designed. This calls for a meticulous design of the hybrid precoder, analog combiners, and scheduling in order to attain ubiquitous multi-group multicast service with low latency. In the following, we present three propositions that will guide our design of $ \left\lbrace \mathcal{V}_{t} \right\rbrace^{T_s}_{t = 1} $, $ \left\lbrace \mathbf{F}_{t} \right\rbrace^{T_s}_{t = 1} $, $ \left\lbrace \mathbf{M}_{t} \right\rbrace^{T_s}_{t = 1} $, $\left\lbrace \mathbf{w}_k \right\rbrace^{K_T}_{k = 1} $ to attain the desired objectives.

\emph{Proposition 1: Let $ 1 \leq t' \leq T_s $ be a window with defined $ \mathcal{V}_{t'} $, $ \mathbf{F}_{t'} $, $ \mathbf{M}_{t'} $, $\left\lbrace \mathbf{w}_k \right\rbrace_{k \in \mathcal{U}_{t'}} $, then the transmission latency $ \xi_{t' \mid \mathcal{V}_{t'}, \mathbf{F}_{t'}, \mathbf{M}_{t'}, \left\lbrace \mathbf{w}_k \right\rbrace_{k \in \mathcal{U}_{t'}}} $ is approximately inversely proportional to the minimum regularized SINR (r-SINR) among all the co-scheduled receivers.}

We assume that for a certain window $ t' $, the sets $ \mathcal{V}_{t'} $ and $ \mathcal{U}_{t'} $ have been determined. Moreover, we consider that $ \mathbf{F}_{t'} $ and $ \mathbf{M}_{t'} $ at the transmitter, and $ \left\lbrace \mathbf{w}_k \right\rbrace_{k \in \mathcal{U}_{t'}} $ for each scheduled receiver, have been designed. Under these assumptions, the SINR at any receiver is obtained via (\ref{e2}). The maximal rate $ \pi_{k \mid \mathcal{V}_{t'}, \mathbf{F}_{t'}, \mathbf{M}_{t'}, \mathbf{w}_k} $ at which receiver $ k \in \mathcal{G}_i, \left( i \in \mathcal{V}_{t'} \right) $ can successfully decode is given by (\ref{e3}). Thus, the minimal latency associated to receiver $ k \in \mathcal{G}_i \left( i \in \mathcal{V}_{t'} \right) $ is $ \xi_{ k \mid \mathcal{V}_{t'}, \mathbf{F}_{t'}, \mathbf{M}_{t'}, \mathbf{w}_k } = \frac{B_i}{\pi_{k \mid \mathcal{V}_{t'}, \mathbf{F}_{t'}, \mathbf{M}_{t'}, \mathbf{w}_k}} $. Since every receiver $ k \in \mathcal{G}_i \left( i \in \mathcal{V}_{t'} \right) $ requires the same information, the optimal rate at which $ \mathbf{b}_i $ can be encoded while guaranteeing successful reception at every receiver, is determined by $ \pi_{i \mid \mathcal{V}_{t'}, \mathbf{F}_{t'}, \mathbf{M}_{t'}, \left\lbrace \mathbf{w}_k \right\rbrace_{k \in \mathcal{U}_{t'}} } = \min_{ k \in \mathcal{G}_i } \pi_{k \mid \mathcal{V}_{t'}, \mathbf{F}_{t'}, \mathbf{M}_{t'}, \mathbf{w}_k } $. The latency associated to multicast group $ i \in \mathcal{V}_{t'} $ is denoted by $ \xi_{ i \mid \mathcal{V}_{t'}, \mathbf{F}_{t'}, \mathbf{M}_{t'}, \left\lbrace \mathbf{w}_k \right\rbrace_{k \in \mathcal{U}_{t'}} } $, and the latency owing to all multicast groups scheduled during window $ t' $ is defined as $ \xi_{t' \mid \mathcal{V}_{t'}, \mathbf{F}_{t'}, \mathbf{M}_{t'}, \left\lbrace \mathbf{w}_k \right\rbrace_{k \in \mathcal{U}_{t'}}} = \max_{ i \in \mathcal{V}_{t'} } \xi_{i \mid \mathcal{V}_{t'}, \mathbf{F}_{t'}, \mathbf{M}_{t'}, \left\lbrace \mathbf{w}_k \right\rbrace_{k \in \mathcal{U}_{t'}} } $, which can also be expressed as
\begin{equation} \label{e4}
	\xi^{\star}_{t'} = \xi_{t' \mid \mathcal{U}_{t'}, \mathbf{F}_{t'}, \mathbf{M}_{t'}, \left\lbrace \mathbf{w}_k \right\rbrace_{k \in \mathcal{U}_{t'}}} = 
							\max_{k \in \mathcal{U}_{t'}} 
								\left\lbrace 
									\frac{B_k}{\pi_{k \mid \mathcal{U}_{t'}, \mathbf{F}_{t'}, \mathbf{M}_{t'}, \mathbf{w}_k}}
								\right\rbrace
\end{equation}

As shown in Appendix \ref{appA}, the transmission latency is approximately inversely proportional to the minimum r-SINR among all the receivers $ k \in \mathcal{U}_{t'} $, i.e.,
\begin{align} \label{e5}
	\begin{split}
		\xi^{\star}_{t'} \approxprop^{-1} 
		\min_{k \in \mathcal{U}_{t'}} 
		\Big\{
				\left( 1 + \beta \cdot \mathrm{SINR}_{k \mid \mathcal{V}_{t'}, \mathbf{F}_{t'}, \mathbf{M}_{t'}, \mathbf{w}_k} \right)^\frac{1}{B_k}
		\Big\},
	\end{split}
\end{align}
where $ (1 + \beta \cdot \mathrm{SINR}_{k \mid \mathcal{V}_{t'}, \mathbf{F}_{t'}, \mathbf{M}_{t'}, \mathbf{w}_k})^\frac{1}{B_k} $ is termed r-SINR. Further, (\ref{e5}) reveals a crucial relation between latency and SINR, which we exploit to design optimal precoders and combiners.

\emph{Proposition 2: Let $ t' $ be a window  with defined $ \mathcal{V}_{t'} $, then the minimization of transmission latency $ \xi_{t' \mid \mathcal{V}_{t'}} $ is equivalent to designing $ \mathbf{F}_{t'}, \mathbf{M}_{t'}, \left\lbrace \mathbf{w}_k \right\rbrace_{k \in \mathcal{U}_{t'}} $ such that the minimum equalized SINR (e-SINR) of all co-scheduled receivers is maximized.}

To minimize the latency associated to $ \mathcal{V}_{t'} $, the optimization problem that devises optimal $ \mathbf{F}_{t'}, \mathbf{M}_{t'}, \left\lbrace \mathbf{w}_k \right\rbrace_{k \in \mathcal{U}_{t'}} $ is
\begin{align} \label{e6}
	\min_{\substack{
					\mathbf{F}_{t'} \in \Omega_F \\
				    \mathbf{M}_{t'} \in \Omega_M \\ 
				    \left\lbrace \mathbf{w}_k \right\rbrace_{k \in \mathcal{U}_{t'}} \in \Omega_W 
			       }
		 } 
		 \xi_{t' \mid \mathcal{U}_{t'}} \leftrightarrow 
	\max_{\substack{
					\mathbf{F}_{t'} \in \Omega_F \\
					\mathbf{M}_{t'} \in \Omega_M \\ 
					\left\lbrace \mathbf{w}_k \right\rbrace_{k \in \mathcal{U}_{t'}} \in \Omega_W
				   }
		 }
		 \min_{k \in \mathcal{U}_{t'}} 
		 	\frac{\mathrm{SINR}_{k \mid \mathcal{V}_{t'}}}{B_k}, 
\end{align}
where $ \frac{\mathrm{SINR}_{k \mid \mathcal{V}_{t'}}}{B_k} $ is e-SINR. Also, $ \Omega_F $, $ \Omega_M $, and $ \Omega_W $ are the feasible sets for the analog precoder, digital precoder and analog combiners, respectively. A more detailed derivation of this expression is available in Appendix \ref{appB}.

\emph{Proposition 3: The minimization of the overall transmission latency of the system is equivalent to simultaneously maximizing the minimum e-SINR of the co-scheduled receivers at every scheduling window}.

It follows from \emph{Proposition 2} that, in order to minimize the overall latency, the latency associated to every scheduling window $ t $ must also be minimum. Thus, 
\begin{align} \label{e7}
	\min_{\substack{
					\left\lbrace \mathbf{F}_t \right\rbrace^{T_s}_{t = 1} \in \Omega_F \\
				    \left\lbrace \mathbf{M}_t \right\rbrace^{T_s}_{t = 1} \in \Omega_M \\ 
				    \left\lbrace \mathbf{w}_k \right\rbrace^{K_T}_{k = 1} \in \Omega_W \\ 
				    \left\lbrace \mathcal{V}_t \right\rbrace^{T_s}_{t = 1} \in \Omega_{V(T_s)} \\ 
				    T_s \in \Omega_{T}
			       }
		 } 
		 \boldsymbol{\xi} \leftrightarrow 
	\max_{\substack{
								\left\lbrace \mathbf{F}_t \right\rbrace^{T_s}_{t = 1} \in \Omega_F \\
	    						\left\lbrace \mathbf{M}_t \right\rbrace^{T_s}_{t = 1} \in \Omega_M \\ 
							    \left\lbrace \mathbf{w}_k \right\rbrace^{K_T}_{k = 1} \in \Omega_W \\ 
							    \left\lbrace \mathcal{V}_t \right\rbrace^{T_s}_{t = 1} \in \Omega_{V(T_s)} \\ 
							    T_s \in \Omega_{T}
				   }
		 }
	\bigwedge^{T_s}_{t = 1} 		 
		 \min_{k \in \mathcal{U}_{t}} 
		 \left\lbrace 
		 	\frac{\mathrm{SINR}_{k \mid \mathcal{V}_{t}}}{B_k} 
		 \right\rbrace,
\end{align}
where $ \boldsymbol{\xi} = \left[ \xi_1, \dots, \xi_{T_s} \right]^T $. Besides, $ \Omega_{V(T_s)} $ denotes the feasible set of all the scheduling combinations for a given $ T_s $, whereas $ \Omega_T \equiv \left[ \ceil[\big]{ G_T / { N^\mathrm{RF}_\mathrm{tx} } }, G_T \right] $ defines the feasible set of $ T_s $. Since the feasible sets $ \Omega_F $, $ \Omega_M $, $ \Omega_W $, $ \Omega_{V(T_s)} $, and $ \Omega_T $ are non-convex (discussed in Section V), (\ref{e7}) is difficult to solve. Moreover, $ \left\lbrace \mathbf{F}_t \right\rbrace^{T_s}_{t = 1} $, $ \left\lbrace \mathbf{M}_t \right\rbrace^{T_s}_{t = 1} $, and $ \left\lbrace \mathbf{w}_k \right\rbrace^{K_T}_{k = 1} $ are mutually coupled as observed in (\ref{e2}). Furthermore, the number of potential scheduling combinations grows combinatorially with $ G_T $ and $ N^\mathrm{RF}_\mathrm{tx} $.

A straightforward but intractable solution to (\ref{e7}) is exhaustive search (\texttt{XHAUS}), whereby every scheduling pattern $ \left\lbrace \mathcal{V}_t \right\rbrace^{T_s}_{t = 1} \in \Omega_{V(T_s)} $ for every value $ T_s \in \Omega_T $ is created. Then, for each pattern, the precoder and combiners are to be designed according to (\ref{e6}) while keeping the best-performing scheduling. A simpler approach consists in randomly associating the groups (\texttt{RAND}) and then solving (\ref{e6}). It is evident that for any scheduling choice, we are required to solve (\ref{e6}) for every window. Thus, in the following, we propose an approach based on semidefinite relaxation and Cholesky matrix factorization to design near-optimal hybrid precoders and analog combiners.

\section{Proposed Joint Design of Hybrid Precoder and Analog Combiners}
Given $ \mathcal{V}_t $, a suitable hybrid precoder and analog combiners for window $ t $ can be designed based on (\ref{e6}).
Thus, the latency $ \xi_t $ is minimized when
\begin{subequations} \label{e8}
	\begin{align}
	\mathcal{P}_0: & \max_{\substack{
													\mathbf{F}_t, \mathbf{M}_t, \\
													\left\lbrace \mathbf{w}_k \right\rbrace_{k \in {\mathcal{U}_t}}
					   				    		  }
			 				  			} 
					& & 			 \min_{k \in {\mathcal{U}_t}}
					 			 		\frac
					 			 		{ \frac{1}{B_k} \left| \mathbf{w}^H_k \mathbf{H}_k \mathbf{F}_t \mathbf{M}_t \mathbf{e}_{i_t} \right|^2}
		 			 					{\displaystyle \sum^{|\mathcal{V}_t|}_{\substack{j_t = 1 \\ j_t \neq i_t }} \left| \mathbf{w}^H_k \mathbf{H}_k \mathbf{F}_t \mathbf{M}_t \mathbf{e}_{j_t} \right|^2 + \sigma^2 \left\| \mathbf{w}_k \right\|^2_2} \label{e8a}
	\\
	\vspace{-0.2cm}
	& ~~~~ \mathrm{s.t.} & & \left\| \mathbf{F}_t \mathbf{M}_t \right\|^2_\mathrm{F} \leq P^\mathrm{max}_\mathrm{tx}, \label{e8b}
	\\
	& & & \left\| \mathbf{w}_k \right\|^2_2 \leq P^\mathrm{max}_\mathrm{rx}, k \in \mathcal{U}_t, \label{e8c}
	\\
	& & & \left[ \mathbf{F}_t \right]_{q,r} \in \mathcal{F},  q \in \mathcal{Q}, r \in \mathcal{R}, \label{e8d}
	\\
	& & & \left[ \mathbf{w}_k \right]_{l} \in \mathcal{W}, l \in \mathcal{L}, \label{e8e}
	\end{align}
\end{subequations}
where (\ref{e8a}) aims at maximizing the minimum e-SINR, (\ref{e8b}) restricts the transmit power of the hybrid precoder, whereas (\ref{e8c}) limits the receive power of each receiver. On the other hand, (\ref{e8d}) and (\ref{e8e}) enforce the phase shifts of the analog precoder $ \mathbf{F}_t $ and analog combiners $ \left\lbrace \mathbf{w}_k \right\rbrace_{k \in {\mathcal{U}_t}} $ to have constant modulus. Furthermore, (\ref{e8a}) is non-convex since it is as a fractional program of quadratic forms with coupled parameters. The constraints (\ref{e8d}) and (\ref{e8e}) are non-convex since $ \left[ \mathbf{F}_t \right]_{q,r} $ and $ \left[ \mathbf{w} \right]_{l} $ belong to the non-convex sets $ \mathcal{F} $ and $ \mathcal{W} $, respectively. Thus, (\ref{e8c}) is also non-convex. Besides, (\ref{e8b}) is non-convex due to the coupling between $ \mathbf{F}_t $ and $ \mathbf{M}_t $, and the existence of (\ref{e8d}). As a result, $ \mathcal{P}_0 $ is a non-convex program with non-convex constrains. Note that (\ref{e8}) can be recast as (\ref{e9}),
\begin{subequations} \label{e9}
	\begin{alignat}{3}
	\mathcal{P}_0: & \max_{\substack{
										\alpha, \mathbf{F}, \mathbf{M}, \\
										\left\lbrace \mathbf{w}_k \right\rbrace^K_{k=1}
					   				}
			 		  	  } & & \alpha \label{e9a} 
	\\
	\vspace{-0.2cm}
	& ~~~ \mathrm{s.t.} & & \frac{1}{B_k} \frac{\left| \mathbf{w}^H_k \mathbf{H}_k \mathbf{F} \mathbf{M} \mathbf{e}_i \right|^2}
		 {\displaystyle \sum_{j \neq i} \left| \mathbf{w}^H_k \mathbf{H}_k \mathbf{F} \mathbf{M} \mathbf{e}_j \right|^2 + \sigma^2 \left\| \mathbf{w}_k \right\|^2_2} \geq \alpha, \label{e9b}
	\\
	& & & \left\| \mathbf{F} \mathbf{M} \right\|^2_\mathrm{F} \leq P^\mathrm{max}_\mathrm{tx}, \label{e9c}
	\\
	& & & \left\| \mathbf{w}_k \right\|^2_2 \leq P^\mathrm{max}_\mathrm{rx}, \forall k \in \mathcal{U}_t, \label{e9d}
	\\
	& & & \left[ \mathbf{F} \right]_{q,r} \in \mathcal{F},  q \in \mathcal{Q}, r \in \mathcal{R}, \label{e9e}
	\\
	& & & \left[ \mathbf{w}_k \right]_{l} \in \mathcal{W}, l \in \mathcal{L}, \label{e9f}
	\\
	& & & \alpha \geq 0. \label{e9g}
	\end{alignat}
\end{subequations}
For notation simplification, we assume that $ G = |\mathcal{V}_t| $ and $ K = |\mathcal{U}_t| $. In addition, we also drop the subscript $ t $ when referring to $ \mathbf{F}_t $ and $ \mathbf{M}_t $. Note that $ \mathcal{P}_0 $ is challenging to solve due to parameter coupling and non-convexity of the feasible sets. In fact, for the particular case when $ \mathbf{F} = \mathbf{I} $ and $ \mathbf{w}_k = 1 $, $ \mathcal{P}_0 $ was shown to be NP-hard \cite{b30}. In this paper, we resort to alternate optimization \cite{b41} to solve (\ref{e9}). Thus, $ \mathbf{F} $, $ \mathbf{M} $, and $ \left\lbrace \mathbf{w}_k \right\rbrace^K_{k=1} $ are sequentially optimized in an iterative manner.

\subsection{Optimization of $\mathbf{F}$}
Given $ \mathbf{M} $ and $ \left\lbrace \mathbf{w}_k \right\rbrace^K_{k=1} $, $ \alpha $ and $ \mathbf{F} $ are optimized as follows,
\begin{subequations} \label{e10}
	\begin{alignat}{3}
	\mathcal{P}_1: & \max_{\alpha, \mathbf{F}} & ~ & \alpha \label{e10a}
	\\
	\vspace{-0.2cm}
	& ~~ \mathrm{s.t.} & & {\alpha B_k \sum_{j \neq i} \left| \mathbf{w}^H_k \mathbf{H}_k \mathbf{F} \mathbf{M} \mathbf{e}_j \right|^2 + \alpha B_k \sigma^2 \left\| \mathbf{w}_k \right\|^2_2} \nonumber
	\\
	& & & - \left| \mathbf{w}^H_k \mathbf{H}_k \mathbf{F} \mathbf{M} \mathbf{e}_i \right|^2 \leq 0, \forall k \in \mathcal{U}_t, \label{e10b}
	\\
	& & & \left\| \mathbf{F} \mathbf{M} \right\|^2_\mathrm{F} \leq P^\mathrm{max}_\mathrm{tx}, \label{e10c}
	\\
	& & & \left[ \mathbf{F} \right]_{q,r} \in \mathcal{F},  q \in \mathcal{Q}, r \in \mathcal{R}, \label{e10d}
	\\
	& & & \alpha \geq 0. \label{e10e}
	\end{alignat}
\end{subequations}

Due to the sub-connected architecture of the analog precoder, $ \mathbf{F} $ is sparse. Thus, we can express $ \mathbf{F} \mathbf{M} = \mathrm{diag} \left( \mathbf{f} \right) \widetilde{\mathbf{M}} $, where $ \widetilde{\mathbf{M}} = \mathbf{M} \otimes \mathbf{1}_{L_{\mathrm{tx}}} $, and $ \mathbf{f} \in \mathbb{C}^{N_\mathrm{tx} \times 1} $ is a vector that contains the non-zero elements of $ \mathbf{F} $. In addition, $ \mathbf{F} \mathbf{M} \mathbf{e}_i = \mathrm{diag} \left( \widetilde{\mathbf{M}}  \mathbf{e}_i \right) \mathbf{f} $. Thus, (\ref{e10}) is equivalent to
\begin{subequations} \label{e11}
	\begin{alignat}{3}
	\mathcal{P}_1: & \max_{\alpha, \mathbf{f}} & ~ & \alpha \label{e11a}
	\\
	\vspace{-0.2cm}
	& ~~ \mathrm{s.t.} & & {\alpha B_k \sum_{j \neq i} \mathbf{f}^H \mathbf{b}_{k,j} \mathbf{b}^H_{k,j} \mathbf{f} - \mathbf{f}^H \mathbf{b}_{k,i} \mathbf{b}^H_{k,i} \mathbf{f}} \nonumber
	\\
	& & & + \alpha B_k \sigma^2 \left\| \mathbf{w}_k \right\|^2_2 \leq 0, \forall k \in \mathcal{U}_t, \label{e11b}
	\\
	& & & \left\| \mathbf{L} \mathbf{f} \right\|^2_2 \leq P^\mathrm{max}_\mathrm{tx}, \label{e11c}
	\\
	& & & \left[ \mathbf{f} \right]_{n} \in \mathcal{F},  n \in \mathcal{N}, \label{e11d}
	\\
	& & & \alpha \geq 0, \label{e11e}
	\end{alignat}
\end{subequations}
where $ \mathbf{b}_{k,i} = \mathrm{diag} \left( \widetilde{\mathbf{M}} \mathbf{e}_i \right)^H \mathbf{H}^H \mathbf{w}^H_k $, $ \mathbf{L} = \left( \widetilde{\mathbf{M}}^T \otimes \mathbf{1}_{N_{\mathrm{tx}}} \right) \odot \left( \mathbf{1}_G \otimes \mathbf{I}_{N_{\mathrm{tx}} \times N_{\mathrm{tx}}} \right) $, $ \mathcal{N} = \left\lbrace 1, 2, \dots, N_{\mathrm{tx}} \right\rbrace $. Realize that (\ref{e11d}) is non-convex due the combinatorial selection of phase shifts $ \left[ \mathbf{f} \right]_n  $ from $ \mathcal{F} $. Similarly, (\ref{e11b}) and (\ref{e11c}) are non-convex as they depend on $ \mathbf{f} $, whereas (\ref{e11e}) is linear. In order to approach $ \mathcal{P}_1 $, our strategy consists of three stages. In the first stage (Stage A\textsubscript{1}), we recast $ \mathcal{P}_1 $ as an SDR program to convexify (\ref{e11b}), (\ref{e11c}), and (\ref{e11d}), thus resulting in $ \mathcal{P}_{\mathrm{SDR},1} $ in (\ref{e12}). After convexification, the only non-convex constraint that remains in $ \mathcal{P}_{\mathrm{SDR},1} $ is (\ref{e12b}). To find a near-optimal solution, we resort to the bisection method in the second stage (Stage A\textsubscript{2}). In the third stage (Stage A\textsubscript{3}), we use Cholesky factorization and randomization to recover $ \mathbf{f} $ from  $ \mathbf{Y} $.


\noindent{\textbf{Stage A\textsubscript{1}} (\textit{Transformation of $ \mathcal{P}_1 $ to SDR}): }
\begin{subequations} \label{e12}
	\begin{alignat}{3}
	\mathcal{P}_{\mathrm{SDR},1}: & \max_{\alpha, \mathbf{Y}} & ~ & \alpha \label{e12a}
	\\
	\vspace{-0.2cm}
	& ~~ \mathrm{s.t.} & & \alpha B_k \mathrm{Tr} \left( \mathbf{B}_{k,\backslash i} \mathbf{Y} \right) - \mathrm{Tr} \left( \mathbf{B}_{k,i} \mathbf{Y} \right) \nonumber 
	\\
	& & & + \alpha B_k \sigma^2 \left\| \mathbf{w}_k \right\|^2_2 \leq 0, \forall k \in \mathcal{U}_t, \label{e12b}
	\\
	& & & \mathrm{Tr} \left( \mathbf{D} \mathbf{Y} \right) \leq P^\mathrm{max}_\mathrm{tx}, \label{e12c}
	\\
	& & & \mathrm{diag} \left( \mathbf{Y} \right) = \delta_F \mathbf{1}_{N_\mathrm{tx}}, \label{e12d}
	\\
	& & & \mathbf{Y} \succcurlyeq 0, \label{e12e}
	\\
	& & & \alpha \geq 0, \label{e12f}
	\end{alignat}
\end{subequations}
where $ \mathbf{Y} = \mathbf{f} \mathbf{f}^H $, $ \mathbf{D} = \mathbf{L}^H \mathbf{L} $, $ \mathbf{B}_{k,\backslash i} = \sum_{j \neq i} \mathbf{b}_{k,j} \mathbf{b}^H_{k,j} $ and $ \mathbf{B}_{k,i} = \mathbf{b}_{k,i} \mathbf{b}^H_{k,i} $ are positive semi-definite matrices. 

\noindent{\textbf{Stage A\textsubscript{2}} (\textit{Bisection search for $ \mathcal{P}_{\mathrm{SDR},1} $}): }
Notice that (\ref{e12b}) is quasi-convex on $ \mathbf{Y} $ and $ \alpha $ because for any $ \alpha \geq 0 $, (\ref{e12b}) collapses to a convex constraint. Thus, we resort to the bisection search method \cite{b5, b25, b30}, where we define an initial interval $ \left[ \alpha^{(0)}_L, \alpha^{(0)}_U \right]  $ for $ \alpha $. Then, we progressively update the interval depending on the returned solutions for $ \mathbf{Y} $. A natural lower bound is $ \alpha^{(0)}_L = 0 $. An upper bound $ \alpha^{(0)}_U $ can be obtained by assigning the total power to the weakest receiver $ k \in \mathcal{U}_t $. Thus, from (\ref{e12b}) we obtain  $ \alpha^{(0)}_U = \min_{k \in \mathcal{U}_t} \frac{\mathrm{Tr} \left( \mathbf{B}_{k,i} \mathbf{Y} \right) }{B_k \sigma^2 \left\| \mathbf{w}_k \right\|^2_2} $. Further, since $ \mathbf{B}_{k,i} $ and $ \mathbf{Y} $ are positive semidefinite, $ \mathrm{Tr} \left( \mathbf{B}_{k,i} \mathbf{Y} \right) \leq \mathrm{Tr} \left( \mathbf{B}_{k,i} \right)  \mathrm{Tr} \left( \mathbf{Y} \right) $ holds \cite{b1}. Also, $ \mathrm{Tr} \left( \mathbf{Y} \right) = \mathrm{Tr} \left( \mathbf{f} \mathbf{f}^H \right) = N^{\mathrm{RF}}_\mathrm{tx} $. Therefore, $ \alpha^{(0)}_U = \min_{k \in \mathcal{U}_t} \frac{ N^{\mathrm{RF}}_\mathrm{tx} \mathrm{Tr} \left( \mathbf{B}_{k,i} \right) }{B_k \sigma^2 \left\| \mathbf{w}_k \right\|^2_2} $. With the initial interval defined, at every iteration $ \ell $ we set $ \alpha^{(\ell)} = \frac{\alpha^{(\ell)}_L + \alpha^{(\ell)}_U}{2} $ and solve $ \mathcal{P}^{(\ell)}_{\mathrm{SDR},1} $ for $ \mathbf{Y} $ with the given $ \alpha^{(\ell)} $. At each iteration $ \ell $, we update the lower and upper bounds in the following manner. If $ \mathcal{P}^{(\ell)}_{\mathrm{SDR},1} $ is feasible, then $ \alpha^{(\ell + 1)}_L = \alpha^{(\ell)} $. Otherwise, $ \alpha^{(\ell + 1)}_U = \alpha^{(\ell)} $. Through this procedure, an $ \epsilon $-suboptimal solution can be obtained within $ N_{\mathrm{bis}_1} = \log_2 \left( \frac{1}{\epsilon} \left( {\alpha^{(0)}_U - \alpha^{(0)}_L} \right) \right) $ iterations.

\noindent{\textbf{Stage A\textsubscript{3}} (\textit{Recovery of $ \mathbf{f} $}): }
Let $ \widehat{\mathbf{Y}} $ represent an $ \epsilon $-suboptimal solution to (\ref{e12}). When $ \widehat{\mathbf{Y}} $ is rank-1, an optimal solution $ \mathbf{Y}^{\star} = \widehat{\mathbf{Y}} $ has been found to (\ref{e12}). Thus, $ \mathbf{f}^{\star} $ can be recovered straightforwardly by means of eigen-decomposition. Otherwise, we resort to a procedure inspired by \cite{b4}. 

\textit{Stage A\textsubscript{3-1}:}
Any element $\left( n_1, n_2 \right) $ of $ \mathbf{Y} $ can be represented as $ \left[ \mathbf{Y} \right]_{n_1,n_2}  = \left[ \mathbf{f} \right]_{n_1} \left[ \mathbf{f} \right]^{*}_{n_2} $. If we define a vector $ \mathbf{u} \in \mathbb{C}^{ N_\mathrm{tx} \times 1} $ such that $ \left\| \mathbf{u}\right\|^2_2 = \mathbf{u}^H \mathbf{u} = 1 $, we can express $ \left[ \mathbf{Y} \right]_{n_1,n_2} $ in terms of $ \mathbf{u} $, i.e., $ \left[ \mathbf{Y} \right]_{n_1,n_2}  = \left( \left[ \mathbf{f} \right]_{n_1} \mathbf{u}^T \right) \left( \left[ \mathbf{f} \right]^{*}_{n_2} \mathbf{u}^{*} \right) $. Moreover, if we define $ \mathbf{q}_{n} = \left[ \mathbf{f} \right]_{n} \mathbf{u} $, $ \mathbf{Y} $ can be recast as $ \mathbf{Y} = \mathbf{Q}^T \mathbf{Q}^{*} $ with $ \mathbf{Q} = \left[ \mathbf{q}_1, \dots, \mathbf{q}_{N_\mathrm{tx}} \right] $. 

\textit{Stage A\textsubscript{3-2}:}
By means of Cholesky factorization, we can decompose $ \widehat{\mathbf{Y}} $ as $ \widehat{\mathbf{Y}} = \widehat{\mathbf{Q}}^T \widehat{\mathbf{Q}}^{*} $, where $ \widehat{\mathbf{Q}} = \left[ \widehat{\mathbf{q}}_1, \dots, \widehat{\mathbf{q}}_{N_\mathrm{tx}} \right]  $. In \textit{Stage A\textsubscript{3-1}}, we assumed that every $ \mathbf{q}_n $ is originated from the same vector $ \mathbf{u} $. Thus, we need to find both $ \mathbf{f} $ and $ \mathbf{u} $ that satisfy $ \mathbf{q}_n = \left[ \mathbf{f} \right]_n \mathbf{u}, \forall n \in \mathcal{N} $. Since such vectors $ \mathbf{f} $ and $ \mathbf{u} $ may not exist, we aim at finding approximate $ \widehat{\mathbf{u}} $  and $ \widehat{\mathbf{f}} $, such that $ \widehat{\mathbf{q}}_n \approx \left[ \widehat{\mathbf{f}} \right]_n \widehat{\mathbf{u}}, \forall n \in \mathcal{N} $. Mathematically, this is expressed as
\begin{subequations} \label{e13}
	\begin{alignat}{3}
	\widehat{\mathcal{P}}_{\mathrm{LS},1}: & \min_{\widehat{\mathbf{u}}, \widehat{\mathbf{f}}} & ~ &
	{ 
			\sum^{N_\mathrm{tx}}_{n = 1} \left\| \widehat{\mathbf{q}}_n - \left[ \widehat{\mathbf{f}} \right]_n  \widehat{\mathbf{u}} \right\|^2_2
	} \label{e13a}
	\\
	\vspace{-0.2cm}
	& \mathrm{s.t.} & & \left\| \widehat{\mathbf{u}} \right\|^2_2 = 1, \label{e13b}
	\\
	& & & \left[ \widehat{\mathbf{f}} \right]_n \in \mathcal{F}, \forall n \in \mathcal{N}, \label{e13c}
	\end{alignat}
\end{subequations}
where we find $ \widehat{\mathbf{u}} $ and $ \widehat{\mathbf{f}} $ with the least error in the 2-norm sense. 

\textit{Stage A\textsubscript{3-3}:} 
Minimizing simultaneously both $ \widehat{\mathbf{u}} $ and $ \widehat{\mathbf{f}} $ is challenging. Thus, we generate a random vector $ \widehat{\mathbf{u}} $, such that $ \left\| \mathbf{u}\right\|^2_2 = 1 $. Therefore, we solve 
\begin{align} \label{e14}
	\widehat{\mathcal{P}}_{\mathrm{LS},2}: & \min_{
												\substack{ 
														\left[ \widehat{\mathbf{f}} \right]_n \in \mathcal{F}, \forall n \in \mathcal{N} \\
													     }  
								        		  }~
	{ 
		\sum^{N_\mathrm{tx}}_{n=1} \left\| \widehat{\mathbf{q}}_n - \left[ \widehat{\mathbf{f}} \right]_n  \widehat{\mathbf{u}} \right\|^2_2
	}
\end{align}

By expanding (\ref{e14}), we obtain
\begin{align} \label{e15}
	\widehat{\mathcal{P}}_{\mathrm{LS},2}: & \max_{
												\substack{ 
															\left[ \widehat{\mathbf{f}} \right]_n \in \mathcal{F}, \forall n \in \mathcal{N} \\
													 }
								   }~
	{ 
		\sum^{N_\mathrm{tx}}_{n=1} \mathfrak{Re} \left( \left[ \widehat{\mathbf{f}} \right]_n \widehat{\mathbf{q}}^H_n \widehat{\mathbf{u}} \right).
	}
\end{align}

Note that (\ref{e15}) can be decomposed into $ N_\mathrm{tx} $ parallel sub-problems. Further, since $ z_n = \widehat{\mathbf{q}}^H_n \widehat{\mathbf{u}} $ is known in $ \widehat{\mathcal{P}}_{\mathrm{LS},2} $, we  select $ \left[ \widehat{\mathbf{f}} \right]_n $ such that the real part is maximized. This is equivalent to choosing $ \left[ \widehat{\mathbf{f}} \right]_n $ with the closest phase to $ z^{*}_n $. Therefore, among the phase rotations in $ \mathcal{F} $, we choose the closest to $ z^{*}_n $. To improve $ \widehat{\mathbf{f}} $,  we generate $ N_{\mathrm{rand}_1} $ vectors $ \widehat{\mathbf{u}}_v $ with $ \left\| \mathbf{u}_v \right\|^2_2 = 1 $ ($ v = 1, \dots, N_{\mathrm{rand}_1} $), and for each we find its corresponding $ \widehat{\mathbf{f}}_v $. Then, we select $ \mathbf{f}^{\dagger} $ among $ N_{\mathrm{rand}_1} $ candidates that provides the largest minimum e-SINR, i.e., $ \mathbf{f}^{\dagger} = \arg_{ \widehat{\mathbf{f}}_1, \dots, \widehat{\mathbf{f}}_{N_{\mathrm{rand}_1}}} \max \min_{k \in \mathcal{U}_t} \frac{\mathrm{SINR}_k}{B_k} $. Finally, $ \mathbf{f}^{\dagger} $  is reshaped to obtain $ \mathbf{F}^{\dagger} = \mathrm{reshape} \left( \mathbf{f}^{\dagger} \right) $. The function $ \mathrm{reshape}(\cdot) $ reverses the effect of $ \mathrm{vec}(\cdot) $.

\subsection{Optimization of $ \mathbf{M} $}
Assuming that $ \mathbf{F} $ and $ \left\lbrace \mathbf{w}_k \right\rbrace^K_{k=1} $ are known, (\ref{e9}) collapses to
\begin{subequations} \label{e16}
	\begin{alignat}{3}
	\mathcal{P}_2: & \max_{\alpha, \mathbf{M}} & ~ & \alpha \label{e16a}
	\\
	\vspace{-0.2cm}
	& ~~ \mathrm{s.t.}& & {\alpha B_k \sum_{j \neq i} \left| \mathbf{w}^H_k \mathbf{H}_k \mathbf{F} \mathbf{M} \mathbf{e}_j \right|^2 + \alpha B_k \sigma^2 \left\| \mathbf{w}_k \right\|^2_2} \nonumber
	\\
	& & & - \left| \mathbf{w}^H_k \mathbf{H}_k \mathbf{F} \mathbf{M} \mathbf{e}_i \right|^2 \leq 0, \forall k \in \mathcal{U}_t, \label{e16b}
	\\
	& & & \left\| \mathbf{F} \mathbf{M} \right\|^2_\mathrm{F} \leq P^\mathrm{max}_\mathrm{tx}, \label{e16c}
	\\
	& & & \alpha \geq 0. \label{e16d}
	\end{alignat}
\end{subequations}

We can equivalently express $ \mathcal{P}_2 $ as,
\begin{subequations} \label{e17}
	\begin{alignat}{3}
	\mathcal{P}_2: & \max_{\alpha, \mathbf{m}} & ~ & \alpha \label{e17a}
	\\
	\vspace{-0.2cm}
	& ~~ \mathrm{s.t.} & & {\alpha B_k \sum_{j \neq i} \mathbf{m}^H \mathbf{c}_{k,j} \mathbf{c}^H_{k,j} \mathbf{m} - \mathbf{m}^H \mathbf{c}_{k,i} \mathbf{c}^H_{k,i} \mathbf{m}}
	\\
	& & & + \alpha B_k \sigma^2 \left\| \mathbf{w}_k \right\|^2_2 \leq 0, \forall k \in \mathcal{U}_t, \label{e17b}
	\\
	& & & \left\| \left( \mathbf{I} \otimes \mathbf{F} \right) \mathbf{m} \right\|^2_2 \leq P^\mathrm{max}_\mathrm{tx}, \label{e17c}
	\\
	& & & \alpha \geq 0, \label{e17d}
	\end{alignat}
\end{subequations}
where $ \mathbf{c}_{k,i} = \left( \mathbf{e}_i \otimes \left( \mathbf{F}^H \mathbf{H}^H_k \mathbf{w}_k \right) \right) $ and $ \mathbf{m} = \mathrm{vec} \left( \mathbf{M} \right) $. Following a similar procedure as before, the SDR form of (\ref{e17}) is
\begin{subequations} \label{e18}
	\begin{align}
	\mathcal{P}_{\mathrm{SDR},2}: & \max_{\alpha, \mathbf{Z}} & ~ & \alpha \label{e18a}
	\\
	\vspace{-0.2cm}
	& ~~ \mathrm{s.t.} & & \alpha B_k \mathrm{Tr} \left( \mathbf{C}_{k,\backslash i} \mathbf{Z} \right) - \mathrm{Tr} \left( \mathbf{C}_{k,i} \mathbf{Z} \right) \nonumber
	\\
	& & & + \alpha B_k \sigma^2 \left\| \mathbf{w}_k \right\|^2_2 \leq 0, \forall k \in \mathcal{U}_t, \quad \quad \quad \label{e18b}
	\\
	& & & \mathrm{Tr} \left( \mathbf{J} \mathbf{Z} \right) \leq P^\mathrm{max}_\mathrm{tx}, \label{e18c}
	\\
	& & & \mathbf{Z} \succcurlyeq 0, \label{e18d}
	\\
	& & & \alpha \geq 0, \label{e18e}
	\end{align}
\end{subequations}
where $ \mathbf{Z} = \mathbf{m} \mathbf{m}^H $, $ \mathbf{J} = \left( \mathbf{I} \otimes \mathbf{F} \right)^H \left( \mathbf{I} \otimes \mathbf{F} \right) $, $ \mathbf{C}_{k,\backslash i} = \sum_{j \neq i} \mathbf{c}_{k,j} \mathbf{c}^H_{k,j} $ and $ \mathbf{C}_{k,i} = \mathbf{c}_{k,i} \mathbf{c}^H_{k,i} $. As in Stage A\textsubscript{2}, we use the bisection method to approach (\ref{e18}). In this case, $ \alpha^{(0)}_L = 0 $ and $ \alpha^{(0)}_U = \min_{k \in \mathcal{U}_t} \frac{ P^{\mathrm{max}}_\mathrm{tx} \mathrm{Tr} \left( \mathbf{C}_{k,i} \right) }{B_k \sigma^2 \left\| \mathbf{w}_k \right\|^2_2} $. The process is repeated for $ N_{\mathrm{bis}_2} $ iterations. At the end of the bisection procedure, we obtain $ \widehat{\mathbf{Z}} $ from which $ \widehat{\mathbf{m}} $ is estimated. If $ \widehat{\mathbf{Z}} $ is rank-1, then an optimal solution $ \mathbf{Z}^{\star} = \widehat{\mathbf{Z}} $ has been found to (\ref{e18}). Otherwise, we generate $ N_{\mathrm{rand}_2} $ candidates according to $ \widehat{\mathbf{m}}_v \sim \mathcal{CN} \left( \mathbf{0}, \widehat{\mathbf{Z}} \right) $ ($ v = 1, \dots, N_{\mathrm{rand}_2} $) and retain the best-performing candidate $ \mathbf{m}^{\dagger} = \arg_{ \widehat{\mathbf{m}}_1, \dots, \widehat{\mathbf{m}}_{N_{\mathrm{rand}_2}}} \max \min_{k \in \mathcal{U}_t} \frac{\mathrm{SINR}_k}{B_k} $ \cite{b2,b3}. Finally, $ \mathbf{m}^{\dagger} $ is reshaped to obtain $ \mathbf{M}^{\dagger} $.

\subsection{Optimization of $ \left\lbrace \mathbf{w}_k \right\rbrace^K_{k = 1} $}
Now, we assume that $ \mathbf{F} $ and $ \mathbf{M} $ are given. Therefore, we optimize the analog combiners $ \left\lbrace \mathbf{w}_k \right\rbrace^K_{k = 1} $ as shown in (\ref{e19})
\begin{subequations} \label{e19}
	\begin{align}
	\mathcal{P}_3: & \max_{\alpha, \left\lbrace \mathbf{w}_k \right\rbrace^K_{k = 1}} & ~ & \alpha \label{e19a}
	\\
	\vspace{-0.2cm}
	& ~~~~~ \mathrm{s.t.} & & {\alpha B_k \sum_{j \neq i} \left| \mathbf{w}^H_k \mathbf{H}_k \mathbf{F} \mathbf{M} \mathbf{e}_j \right|^2 + \alpha B_k \sigma^2 \left\| \mathbf{w}_k \right\|^2_2} \nonumber
	\\
	& & & - \left| \mathbf{w}^H_k \mathbf{H}_k \mathbf{F} \mathbf{M} \mathbf{e}_i \right|^2 \leq 0, \forall k \in \mathcal{U}_t, \label{e19b}
	\\
	& & & \left\| \mathbf{w}_k \right\|^2_2 \leq P^\mathrm{max}_\mathrm{rx}, \forall k \in \mathcal{U}_t, \label{e19c}
	\\
	& & & \left[ \mathbf{w}_k \right]_{l} \in \mathcal{W}, l \in \mathcal{L}, \label{e19d}
	\\
	& & & \alpha \geq 0. \label{e19e}
	\end{align}
\end{subequations}

We observe that each combiner can be optimized independently since any variation of $ \mathbf{w}_k $ will only affect the SINR of the $ k $-th receiver. Therefore, we solve $ K $ sub-problems $ \mathcal{P}^{(k)}_{3} $ in parallel. The SDR form of $ \mathcal{P}^{(k)}_{3} $ is
\begin{subequations} \label{e20}
	\begin{align}
	\mathcal{P}^{(k)}_{\mathrm{SDR},3}: & \max_{\alpha, \mathbf{W}_k} & ~ & \alpha \label{e20a}
	\\
	\vspace{-0.2cm}
	& ~~ \mathrm{s.t.} & & \alpha B_k \mathrm{Tr} \left( \widetilde{\mathbf{P}}_{k,\backslash i} \mathbf{W}_k \right) - \mathrm{Tr} \left( \mathbf{P}_{k,i} \mathbf{W}_k \right), \label{e20b}
	\\
	& & & \mathrm{Tr} \left( \mathbf{W}_k \right) \leq P^\mathrm{max}_\mathrm{rx}, \label{e20c}
	\\
	& & & \mathrm{diag} \left( \mathbf{W}_k \right) = \delta_W \mathbf{1}_{N_\mathrm{rx}}, \label{e20d}
	\\
	& & & \mathbf{W}_k \succcurlyeq 0, \label{e20e}
	\\
	& & & \alpha \geq 0, \label{e20f}
	\end{align}
\end{subequations}
where $ \mathbf{W}_k = \mathbf{w}_k \mathbf{w}^H_k $, $ \mathbf{P}_{k,i} = \mathbf{H}_k \mathbf{F} \mathbf{M} \mathbf{e}_i \mathbf{e}^H_i \mathbf{M}^H \mathbf{F}^H \mathbf{H}_k $ and $ \widetilde{\mathbf{P}}_{k,\backslash i} = \sum_{j \neq i} \mathbf{P}_{k,j} + B_k \sigma^2 \mathbf{I} $. Since the combiners $ \mathbf{w}_k $ are analog, we follow the same procedure used to optimize $ \mathbf{F} $. In this case, the lower bound is $ \alpha^{(0)}_{L_k} = 0 $ whereas the upper bound $ \alpha^{(0)}_{U_k} = \lambda_{\mathrm{max}} \left( \mathbf{P}_{k,i} \widetilde{\mathbf{P}}^{-1}_{k,\backslash i} \right) $ is the maximum eigenvalue of matrix $ \mathbf{P}_{k,i} \widetilde{\mathbf{P}}^{-1}_{k,\backslash i} $. The bisection process is repeated for $ N_{\mathrm{bis}_3} $ iterations as explained in Stage A\textsubscript{2}, upon whose completion $ \widehat{\mathbf{W}}_k $ is obtained. If $ \widehat{\mathbf{W}}_k $ is rank-1, then $ \mathbf{W}^{\star}_k = \widehat{\mathbf{W}}_k $ is also optimal to (\ref{e20}). Otherwise we generate $N_{\mathrm{rand}_3} $ candidates $ \widehat{\mathbf{w}}_{k,v} $ ($ v = 1, \dots, N_{\mathrm{rand}_3} $) for each $ \widehat{\mathbf{w}}_k $ as discussed in Stage A\textsubscript{3}, and select the best-performing $ \mathbf{w}^{\dagger}_k = \arg_{ \widehat{\mathbf{w}}_{k,1}, \dots, \widehat{\mathbf{w}}_{k,N_{\mathrm{rand}_3}}} \max \frac{\mathrm{SINR}_k}{B_k} $. 

To further refine $ \mathbf{F} $, $ \mathbf{M} $, $ \left\lbrace \mathbf{w}_k \right\rbrace^K_{k = 1} $, we sequentially solve $ \mathcal{P}_{\mathrm{SDR},1} $, $ \mathcal{P}_{\mathrm{SDR},2} $ and $ \left\lbrace \mathcal{P}^{(k)}_{\mathrm{SDR},3} \right\rbrace^{K}_{k = 1} $ for a number of $ N_\mathrm{iter} $ iterations. Also, since the number of bits $ B_i $ to be transmitted can be arbitrarily large, we use normalized values $ \widetilde{B}_i $ in the range $ \left[ 0, 1 \right] $, only for optimization purposes. Also, the scaling factors $ \delta_F $ and $ \delta_W $, in Section III, are chosen such that $ \left\| \mathbf{F} \right\|^2_\mathrm{F} = N^\mathrm{RF}_\mathrm{tx} $ and $ \left\| \mathbf{w}_k \right\|^2_2 = P^\mathrm{max}_\mathrm{rx} $. Thus, $ \delta_F = N^\mathrm{RF}_\mathrm{tx} / N_\mathrm{tx} $ and $ \delta_W = P^\mathrm{max}_\mathrm{rx} / N_\mathrm{rx} $.

\section{Proposed Scheduling Algorithm}
\label{sec:scheduling_algo}

Given that scheduling plays a key role in ensuring minimum latency, in this section, we propose a novel scheduling formulation that aims to minimize both, the number of scheduling windows and the aggregate inter-group correlation (IGC). We model the scheduling problem as a Boolean program,
\begin{subequations} \label{e21}
	\begin{align}
	\mathcal{S}: & \min_{ \left\lbrace \mu_{i,i} \right\rbrace^{G_T}_{i = 1}, \left\lbrace \tau_{i,j,l} \right\rbrace } & & \underbrace{ \sum^{G_T}_{i = 1} \mu_{i,i} }_{\text{first term}} + \omega \underbrace { \sum^{G_T - 1}_{i = 1} \sum^{G_T - 1}_{j \geq i} \sum^{G_T}_{l > j} \rho_{j,l} \cdot \tau_{i,j,l} }_{\text{second term: aggregate IGC}} \label{e21a}
	\\
	\vspace{-0.2cm}
	& ~~~~ \mathrm{s.t.} & & \sum_{ i \leq j < l} \rho_{j,l} \cdot  \tau_{i,j,l} \leq \lambda \cdot \mu_{i,i}, \forall i, \label{e21b}
	\\
	& & & \sum_{i \leq j} \mu_{i,j} = 1, \forall j, \label{e21c}	
	\\
	& & & \sum_{j \geq i} \mu_{i,j} \leq N^{\mathrm{RF}}_{\mathrm{tx}}, \forall i, \label{e21d}	
	\\
	& & & \mu_{i,j} \leq \mu_{i,i}, \forall i < j, \label{e21e}
	\\
	& & & \mu_{i,j} + \mu_{i,l} \leq 1 + \tau_{i,j,l}, \forall i \le j < l, \label{e21f}
	\\
	& & & \mu_{i,j} \in \left\lbrace 0, 1 \right\rbrace, \label{e21g}
	\\
	& & & \tau_{i,j,l} \in \left\lbrace 0, 1 \right\rbrace. \label{e21h}
	\end{align}
\end{subequations}

Each binary variable $ \mu_{i,j} $ assumes the value of $ 1 $ if multicast group $ j $ is scheduled in the $ i $-th window (or $ 0 $ otherwise). The binary variable $ \tau_{i,j,l} $ is $ 1 $ if any two multicast groups $ j $ and $ l $ have been co-scheduled during the $ i $-th window. Also, we define $ \rho_{j,l} = \frac{ \left| \mathbf{h}^H_j \mathbf{h}_l \right| }{ \left\| \mathbf{h}_j \right\|_2 \left\| \mathbf{h}_l \right\|_2 } $ as the inter-group correlation (IGC) between groups $ j $ and $ l $, where $ \mathbf{h}_j = \frac{1}{|\mathcal{G}_j|} \sum_{k \in \mathcal{G}_j} \mathrm{vec} \left( \mathbf{H}_k \right) $ is the mean channel vector of all receivers $ k $ in group $ j $. The first term in (\ref{e21a}) represents the number of scheduling windows, whereas the second term is the aggregate IGC, which is computed in a pair-wise manner and accumulated as a penalization. If there exist more than a solution that yields the same optimal amount of scheduling windows, the second term penalizes the candidates that exhibit large aggregate IGC. Realize that by minimizing the aggregate IGC, we attempt to produce scheduling patterns wherein receivers of different co-scheduled groups are the least correlated (on average), thereby enhancing the SINR and latency. For every window $ i $, (\ref{e21b}) restricts the aggregate IGC of the co-scheduled groups to remain below $ \lambda $, which is the maximum threshold. Also, (\ref{e21c}) enforces every group $ j $ to be scheduled once, whereas (\ref{e21d}) restricts the number of groups per window to be at most $ N^{\mathrm{RF}}_{\mathrm{tx}} $. Without loss of generality, (\ref{e21e}) and the condition $ i < j $ imposed on (\ref{e21a})--(\ref{e21d}) reduce the search space and therefore the complexity. Further, (\ref{e21f}) binds the variables $ \mu_{i,j} $ and $ \tau_{i,j,l} $ and ensures consistency among them. Finally, (\ref{e21g}) and (\ref{e21h}) declare $ \mu_{i,i} $ and $ \tau_{i,j,l} $ as Boolean. In (\ref{e21a}), $ \omega $ is chosen such that the first and second term have equal weights on average.

\section{Numerical Results}
This section sheds light on the performance of our proposed scheme \texttt{\small{HYDRAWAVE}} (joint group scheduling and precoding). We focus on two performance metrics: \emph{minimum e-SINR} and \emph{latency}. First, to gain insights into the effectiveness of our hybrid precoder design, we evaluate its performance against fully-digital and fully-analog implementations, in which we leave out the scheduling aspect. Next, we investigate the impact of different  scheduling algorithms on the total latency, which additionally accounts for the beam-switching delay between the scheduling windows.\\
\noindent{{\bf\texttt{\small{HYDRAWAVE}}:} our proposed scheduling and precoding scheme.} \\
\noindent{{\bf\texttt{\small{SING}}:} single-group multicasting scheduling serves only one multicast group per scheduling window.} \\
\noindent{{\bf\texttt{\small{RAND}}:} random scheduling selects stochastically an allocation pattern among all possible combinations.} \\
\noindent{{\bf\texttt{\small{XHAUS}}:} exhaustive search finds the best scheduling policy that minimizes latency among all the possibilities.} 
%
%
%

In simulations, we consider the geometric channel model with $ N_{\mathrm{paths}} = 6 $ propagation paths between the transmitter and each receiver \cite{b13, b16}, considering the high density of reflecting surfaces in an industrial environment. The receivers are divided equally among all the multicast groups. The numerical results show the average performance over $100$ channel realizations. \tref{tab:param} summarizes the parameters setting.

\begin{figure*}[t] \nonumber
	\begin{minipage}{0.9\textwidth}
	\centering
	\begin{subfigure}[t]{0.32\textwidth}
		\begin{tikzpicture}[inner sep = 0.3mm]
		\begin{axis}[
		ybar,
		ymin = 0,
		ymax = 175,
		width = 6.0cm,
		height = 3.0cm,
		bar width = 6pt,
		tick align = inside,
		x label style={align=center, font=\footnotesize,},
		ylabel = {Min. e-SINR},
		y label style={at={(-0.1,0.5)}, font=\footnotesize,},
		symbolic x coords = {Nrx1, Nrx2, Nrx3},
		xticklabels = {$ N_\mathrm{rx} = 1 $, $ N_\mathrm{rx} = 2 $, $ N_\mathrm{rx} = 3 $},
		x tick label style = {text width = 2cm, align = center, font = \fontsize{7}{8}\selectfont},
		y tick label style = {text width = 0.8cm, align = right, font = \fontsize{7}{8}\selectfont},
		xtick = data,
		enlarge y limits = {value = 0.3, upper},
		enlarge x limits = 0.24,
		legend columns = 5,
		legend pos = north east,
		legend style={at={(0.6,1.10)},anchor=south west, font=\fontsize{6}{5}\selectfont, text width=1.55cm,text height=0.02cm,text depth=.ex, fill = none},
		nodes near coords bottom/.style={
    		scatter/position=absolute,
    		close to zero/.style={
        		at={(axis cs:\pgfkeysvalueof{/data point/x},\pgfkeysvalueof{/data point/y})},
    		},
    		big value/.style={
        		at={(axis cs:\pgfkeysvalueof{/data point/x},\pgfkeysvalueof{/data point/y})},
        		color = black, text opacity=1, 
        		inner ysep = 0.5pt, yshift = -21pt 
    		},
    		every node near coord/.append style={
      		check for zero/.code={%
        		\pgfmathfloatifflags{\pgfplotspointmeta}{0}{%
            		\pgfkeys{/tikz/coordinate}%
        		}{%
            		\begingroup
            		\pgfkeys{/pgf/fpu}%
            		\pgfmathparse{\pgfplotspointmeta<#1}%
            		\global\let\result=\pgfmathresult
            		\endgroup
            		\pgfmathfloatcreate{1}{1.0}{0}%
            		\let\ONE=\pgfmathresult
            		\ifx\result\ONE
                		\pgfkeysalso{/pgfplots/close to zero}%
            		\else
                		\pgfkeysalso{/pgfplots/big value}%
            		\fi
        		}
      		},
      		check for zero, 
      		rotate=90, anchor = west, font = \fontsize{0.5}{1}\selectfont,
    		},%
		},%
	nodes near coords,
	nodes near coords={\pgfmathprintnumber[fixed zerofill,precision=2]{\pgfplotspointmeta}},
	nodes near coords bottom = 105,  
		]
		
		\addplot[fill = bcolor1] coordinates {(Nrx1, 101.0345) (Nrx2, 167.5321) (Nrx3, 216.0149)};
		
		\addplot[fill = bcolor2] coordinates {(Nrx1, 60.3298) (Nrx2, 99.9330) (Nrx3, 129.4011)};
		
		\addplot[fill = bcolor3] coordinates {(Nrx1, 36.3392) (Nrx2, 62.2636) (Nrx3, 79.0546)};
		
		\addplot[fill = bcolor4] coordinates {(Nrx1, 18.8461) (Nrx2, 34.1564) (Nrx3, 46.3972)};
		
		\addplot[fill = bcolor5] coordinates {(Nrx1, 8.5153) (Nrx2, 17.0937) (Nrx3, 25.0711)};
	
		\end{axis}
		\end{tikzpicture}
	\end{subfigure}%
	\hspace{3mm}
	\begin{subfigure}[t]{0.32\textwidth}
		\begin{tikzpicture}[inner sep = 0.3mm]
		\begin{axis}[
		ybar,
		ymin = 0,
		ymax = 75,
		width = 6.0cm,
		height = 3.0cm,
		bar width = 6pt,
		tick align = inside,
		x label style={align=center, font=\footnotesize,},
		y label style={at={(-0.1,0.5)}, font=\footnotesize,},
		symbolic x coords = {Nrx1, Nrx2, Nrx3},
		xticklabels = {$ N_\mathrm{rx} = 1 $, $ N_\mathrm{rx} = 2 $, $ N_\mathrm{rx} = 3 $},
		x tick label style = {text width = 2cm, align = center, font = \fontsize{7}{8}\selectfont},
		y tick label style = {text width = 0.5cm, align = right, font = \fontsize{7}{8}\selectfont},
		xtick = data,
		enlarge y limits = {value = 0.4, upper},
		enlarge x limits = 0.24,
		nodes near coords bottom/.style={
    		scatter/position=absolute,
    		close to zero/.style={
        		at={(axis cs:\pgfkeysvalueof{/data point/x},\pgfkeysvalueof{/data point/y})},
    		},
    		big value/.style={
        		at={(axis cs:\pgfkeysvalueof{/data point/x},\pgfkeysvalueof{/data point/y})},
        		color = black, text opacity=1, 
        		inner ysep = 0.5pt, yshift = -18pt 
    		},
    		every node near coord/.append style={
      		check for zero/.code={%
        		\pgfmathfloatifflags{\pgfplotspointmeta}{0}{%
            		\pgfkeys{/tikz/coordinate}%
        		}{%
            		\begingroup
            		\pgfkeys{/pgf/fpu}%
            		\pgfmathparse{\pgfplotspointmeta<#1}%
            		\global\let\result=\pgfmathresult
            		\endgroup
            		\pgfmathfloatcreate{1}{1.0}{0}%
            		\let\ONE=\pgfmathresult
            		\ifx\result\ONE
                		\pgfkeysalso{/pgfplots/close to zero}%
            		\else
                		\pgfkeysalso{/pgfplots/big value}%
            		\fi
        		}
      		},
      		check for zero, 
      		rotate=90, anchor = west, font = \fontsize{0.5}{1}\selectfont,
    		},%
		},%
	nodes near coords,
	nodes near coords={\pgfmathprintnumber[fixed zerofill,precision=2]{\pgfplotspointmeta}},
	nodes near coords bottom = 70,  
		]
		
		\addplot[fill = bcolor1] coordinates {(Nrx1, 47.6338) (Nrx2, 76.3093) (Nrx3, 98.4381)};
		
		\addplot[fill = bcolor2] coordinates {(Nrx1, 24.4036) (Nrx2, 39.1340) (Nrx3, 48.8037)};
				
		\addplot[fill = bcolor3] coordinates {(Nrx1, 10.5616) (Nrx2, 18.6608) (Nrx3, 23.3403)};
		
		\addplot[fill = bcolor4] coordinates {(Nrx1, 4.6031) (Nrx2, 8.3678) (Nrx3, 10.7263)};
		
		\addplot[fill = bcolor5] coordinates {(Nrx1, 2.1031) (Nrx2, 4.0046) (Nrx3, 5.2668)};
		
		\end{axis}
		\end{tikzpicture}
	\end{subfigure}%
	\hspace{1mm}
	\begin{subfigure}[t]{0.32\textwidth}	
		\centering
		\begin{tikzpicture}[inner sep = 0.3mm]
		\begin{axis}[
		ybar,
		ymin = 0,
		ymax = 31,
		width = 6.0cm,
		height = 3.0cm,
		bar width = 6pt,
		tick align = inside,
		x label style={align=center, font=\footnotesize,},
		y label style={at={(-0.1,0.5)}, font=\footnotesize,},
		symbolic x coords = {Nrx1, Nrx2, Nrx3},
		xticklabels = {$ N_\mathrm{rx} = 1 $, $ N_\mathrm{rx} = 2 $, $ N_\mathrm{rx} = 3 $},
		x tick label style = {text width = 2cm, align = center, font = \fontsize{7}{8}\selectfont},
		y tick label style = {text width = 0.5cm, align = right, font = \fontsize{7}{8}\selectfont},
		xtick = data,
		enlarge y limits = {value = 0.2, upper},
		enlarge x limits = 0.24,
		legend columns = 1,
		legend pos = north east,
		legend style={at={(1,0)},anchor=south west, font=\fontsize{6}{5}\selectfont, text width=1.0cm,text height=0.02cm,text depth=.ex, fill = none},
		nodes near coords bottom/.style={
    		scatter/position=absolute,
    		close to zero/.style={
        		at={(axis cs:\pgfkeysvalueof{/data point/x},\pgfkeysvalueof{/data point/y})},
    		},
    		big value/.style={
        		at={(axis cs:\pgfkeysvalueof{/data point/x},\pgfkeysvalueof{/data point/y})},
        		color = black, text opacity=1, 
        		inner ysep = 0.5pt, yshift = -18pt 
    		},
    		every node near coord/.append style={
      		check for zero/.code={%
        		\pgfmathfloatifflags{\pgfplotspointmeta}{0}{%
            		\pgfkeys{/tikz/coordinate}%
        		}{%
            		\begingroup
            		\pgfkeys{/pgf/fpu}%
            		\pgfmathparse{\pgfplotspointmeta<#1}%
            		\global\let\result=\pgfmathresult
            		\endgroup
            		\pgfmathfloatcreate{1}{1.0}{0}%
            		\let\ONE=\pgfmathresult
            		\ifx\result\ONE
                		\pgfkeysalso{/pgfplots/close to zero}%
            		\else
                		\pgfkeysalso{/pgfplots/big value}%
            		\fi
        		}
      		},
      		check for zero, 
      		rotate=90, anchor = west, font = \fontsize{0.5}{1}\selectfont,
    		},%
		},%
	nodes near coords,
	nodes near coords={\pgfmathprintnumber[fixed zerofill,precision=2]{\pgfplotspointmeta}},
	nodes near coords bottom = 20,  
		]

		\addplot[fill = bcolor1] coordinates {(Nrx1, 15.9837) (Nrx2, 26.3660) (Nrx3, 35.5727)}; 
		
		\addplot[fill = bcolor2] coordinates {(Nrx1, 8.8593) (Nrx2, 14.9988) (Nrx3, 20.2280)}; 
			
		\addplot[fill = bcolor3] coordinates {(Nrx1, 4.3380) (Nrx2, 7.5459) (Nrx3, 10.3651)}; 
		
		\addplot[fill = bcolor4] coordinates {(Nrx1, 2.1725) (Nrx2, 3.9129) (Nrx3, 5.5151)}; 
				
		\addplot[fill = bcolor5] coordinates {(Nrx1, 0.9574) (Nrx2, 1.9910) (Nrx3, 2.9776)}; 
		
		\end{axis}
		\end{tikzpicture}
	\end{subfigure}
	\begin{subfigure}[t]{0.32\textwidth}
		\begin{tikzpicture}[inner sep = 0.3mm]
		\begin{axis}
		[
		ybar,
		ymin = 0,
		ymax = 1.4,
		width = 6.0cm,
		height = 3.0cm,
		bar width = 6pt,
		tick align = inside,
		x label style={align=center, font=\footnotesize,},
		ylabel = {Latency (ms)},
		y label style={at={(-0.1,0.5)}, font=\footnotesize,},
		symbolic x coords = {Nrx1, Nrx2, Nrx3},
		xticklabels = {$ N_\mathrm{rx} = 1 $, $ N_\mathrm{rx} = 2 $, $ N_\mathrm{rx} = 3 $},
		x tick label style = {text width = 2cm, align = center, font = \fontsize{7}{8}\selectfont},
		y tick label style = {text width = 0.8cm, align = right, font = \fontsize{7}{8}\selectfont},
		xtick = data,
		enlarge y limits = {value = 0.3, upper},
		enlarge x limits = 0.24,
		legend columns = 8,
		legend pos = north east,
		legend style={at={(0.015,0.78)},anchor=south west, font=\fontsize{6}{5}\selectfont, text width=1.55cm,text height=0.02cm,text depth=.ex, fill = none},
		nodes near coords bottom/.style={
    		scatter/position=absolute,
    		close to zero/.style={
        		at={(axis cs:\pgfkeysvalueof{/data point/x},\pgfkeysvalueof{/data point/y})},
    		},
    		big value/.style={
        		at={(axis cs:\pgfkeysvalueof{/data point/x},\pgfkeysvalueof{/data point/y})},
        		color = black, text opacity=1, 
        		inner ysep = 0.5pt, yshift = -16pt 
    		},
    		every node near coord/.append style={
      		check for zero/.code={%
        		\pgfmathfloatifflags{\pgfplotspointmeta}{0}{%
            		\pgfkeys{/tikz/coordinate}%
        		}{%
            		\begingroup
            		\pgfkeys{/pgf/fpu}%
            		\pgfmathparse{\pgfplotspointmeta<#1}%
            		\global\let\result=\pgfmathresult
            		\endgroup
            		\pgfmathfloatcreate{1}{1.0}{0}%
            		\let\ONE=\pgfmathresult
            		\ifx\result\ONE
                		\pgfkeysalso{/pgfplots/close to zero}%
            		\else
                		\pgfkeysalso{/pgfplots/big value}%
            		\fi
        		}
      		},
      		check for zero, 
      		rotate=90, anchor = west, font = \fontsize{0.5}{1}\selectfont,
    		},%
		},%
		nodes near coords,
		nodes near coords={\pgfmathprintnumber[fixed zerofill,precision=2]{\pgfplotspointmeta}},
		nodes near coords bottom = 0.8,  
		]
		
		\addplot[fill = bcolor1] coordinates {(Nrx1, 0.8364) (Nrx2, 0.8334) (Nrx3, 0.8333)};
		
		\addplot[fill = bcolor2] coordinates {(Nrx1, 0.8549) (Nrx2, 0.8365) (Nrx3, 0.8336)};
		
		\addplot[fill = bcolor3] coordinates {(Nrx1, 0.9562) (Nrx2, 0.8503) (Nrx3, 0.8411)};
		
		\addplot[fill = bcolor4] coordinates {(Nrx1, 1.2099) (Nrx2, 0.9777) (Nrx3, 0.8947)};
		
		\addplot[fill = bcolor5, every node near coord/.append style={color = black}] coordinates {(Nrx1, 1.7379) (Nrx2, 1.2765) (Nrx3, 1.0887)};
	
		\end{axis}
		\end{tikzpicture}
		\vspace{-6mm}
		\subcaption{Fully-digital precoder ($ N_{\mathrm{tx}}^{\mathrm{RF}} = 24 $)}
	\end{subfigure}%
	~
	\begin{subfigure}[t]{0.32\textwidth}
		\hspace{1mm}
		\begin{tikzpicture}[inner sep = 0.3mm]
		\begin{axis}[
		ybar,
		ymin = 0,
		ymax = 3.25,
		width = 6.0cm,
		height = 3.0cm,
		bar width = 6pt,
		tick align = inside,
		x label style={align=center, font=\footnotesize,},
		y label style={at={(-0.1,0.5)}, font=\footnotesize,},
		symbolic x coords = {Nrx1, Nrx2, Nrx3},
		xticklabels = {$ N_\mathrm{rx} = 1 $, $ N_\mathrm{rx} = 2 $, $ N_\mathrm{rx} = 3 $},
		x tick label style = {text width = 2cm, align = center, font = \fontsize{7}{8}\selectfont},
		y tick label style = {text width = 0.5cm, align = right, font = \fontsize{7}{8}\selectfont},
		xtick = data,
		enlarge y limits = {value = 0.4, upper},
		enlarge x limits = 0.24,
		legend columns = 8,
		legend pos = north east,
		legend style={at={(0.015,0.78)},anchor=south west, font=\fontsize{6}{5}\selectfont, text width=1.55cm,text height=0.02cm,text depth=.ex, fill = none},
		nodes near coords bottom/.style={
    		scatter/position=absolute,
    		close to zero/.style={
        		at={(axis cs:\pgfkeysvalueof{/data point/x},\pgfkeysvalueof{/data point/y})},
    		},
    		big value/.style={
        		at={(axis cs:\pgfkeysvalueof{/data point/x},\pgfkeysvalueof{/data point/y})},
        		color = black, text opacity=1, 
        		inner ysep = 0.5pt, yshift = -16pt 
    		},
    		every node near coord/.append style={
      		check for zero/.code={%
        		\pgfmathfloatifflags{\pgfplotspointmeta}{0}{%
            		\pgfkeys{/tikz/coordinate}%
        		}{%
            		\begingroup
            		\pgfkeys{/pgf/fpu}%
            		\pgfmathparse{\pgfplotspointmeta<#1}%
            		\global\let\result=\pgfmathresult
            		\endgroup
            		\pgfmathfloatcreate{1}{1.0}{0}%
            		\let\ONE=\pgfmathresult
            		\ifx\result\ONE
                		\pgfkeysalso{/pgfplots/close to zero}%
            		\else
                		\pgfkeysalso{/pgfplots/big value}%
            		\fi
        		}
      		},
      		check for zero, 
      		rotate=90, anchor = west, font = \fontsize{0.5}{1}\selectfont,
    		},%
		},%
	nodes near coords,
	nodes near coords={\pgfmathprintnumber[fixed zerofill,precision=2]{\pgfplotspointmeta}},
	nodes near coords bottom = 2,  
		]
		
		\addplot[fill = bcolor1] coordinates {(Nrx1, 0.9158) (Nrx2, 0.8608) (Nrx3, 0.8426)};
		
		\addplot[fill = bcolor2] coordinates {(Nrx1, 1.1108) (Nrx2, 0.9444) (Nrx3, 0.8993)};
				
		\addplot[fill = bcolor3] coordinates {(Nrx1, 1.5821) (Nrx2, 1.2325) (Nrx3, 1.1277)};
		
		\addplot[fill = bcolor4] coordinates {(Nrx1, 2.5208) (Nrx2, 1.7824) (Nrx3, 1.5804)};
		
		\addplot[fill = bcolor5, every node near coord/.append style={color = black}] coordinates {(Nrx1, 4.1733) (Nrx2, 2.6996) (Nrx3, 2.2955)};
		
		\end{axis}
		\end{tikzpicture}
		\vspace{-6mm}
		\subcaption{Hybrid precoder ($ N_{\mathrm{tx}}^{\mathrm{RF}} = 4 $) }
	\end{subfigure}%
	~
	\begin{subfigure}[t]{0.32\textwidth}	
		\hspace{0.5mm}
		\begin{tikzpicture}[inner sep = 0.3mm]
		\begin{axis}[
		ybar,
		ymin = 0,
		ymax = 7,
		width = 6.0cm,
		height = 3.0cm,
		bar width = 6pt,
		tick align = inside,
		x label style={align=center, font=\footnotesize,},
		y label style={at={(-0.1,0.5)}, font=\footnotesize,},
		symbolic x coords = {Nrx1, Nrx2, Nrx3},
		xticklabels = {$ N_\mathrm{rx} = 1 $, $ N_\mathrm{rx} = 2 $, $ N_\mathrm{rx} = 3 $},
		x tick label style = {text width = 2cm, align = center, font = \fontsize{7}{8}\selectfont},
		y tick label style = {text width = 0.5cm, align = right, font = \fontsize{7}{8}\selectfont},
		xtick = data,
		enlarge y limits = {value = 0.2, upper},
		enlarge x limits = 0.24,
		legend columns = 5,
		legend pos = north east,
		legend style={at={(-0.08,-0.88)},anchor=south east, font=\fontsize{6}{5}\selectfont, text width=1.15cm,text height=0.02cm, text depth=.ex, fill = none},
		nodes near coords bottom/.style={
    		scatter/position=absolute,
    		close to zero/.style={
        		at={(axis cs:\pgfkeysvalueof{/data point/x},\pgfkeysvalueof{/data point/y})},
    		},
    		big value/.style={
        		at={(axis cs:\pgfkeysvalueof{/data point/x},\pgfkeysvalueof{/data point/y})},
        		inner ysep = 0.5pt, yshift = -16pt 
    		},
    		every node near coord/.append style={
      		check for zero/.code={%
        		\pgfmathfloatifflags{\pgfplotspointmeta}{0}{%
            		\pgfkeys{/tikz/coordinate}%
        		}{%
            		\begingroup
            		\pgfkeys{/pgf/fpu}%
            		\pgfmathparse{\pgfplotspointmeta<#1}%
            		\global\let\result=\pgfmathresult
            		\endgroup
            		\pgfmathfloatcreate{1}{1.0}{0}%
            		\let\ONE=\pgfmathresult
            		\ifx\result\ONE
                		\pgfkeysalso{/pgfplots/close to zero}%
            		\else
                		\pgfkeysalso{/pgfplots/big value}%
            		\fi
        		}
      		},
      		check for zero, 
      		rotate=90, anchor = west, font = \fontsize{0.5}{1}\selectfont,
    		},%
		},%
	nodes near coords,
	nodes near coords={\pgfmathprintnumber[fixed zerofill,precision=2]{\pgfplotspointmeta}},
	nodes near coords bottom = 4,  
		]
		
		\addplot[fill = bcolor1] coordinates {(Nrx1, 1.4768) (Nrx2, 1.1572) (Nrx3, 1.0544)};

		\addplot[fill = bcolor2] coordinates {(Nrx1, 1.8373) (Nrx2, 1.4284) (Nrx3, 1.2545)};

		\addplot[fill = bcolor3] coordinates {(Nrx1, 2.6733) (Nrx2, 1.9092) (Nrx3, 1.6104)};

		\addplot[fill = bcolor4] coordinates {(Nrx1, 4.2911) (Nrx2, 2.8347) (Nrx3, 2.2721)};

		\addplot[fill = bcolor5] coordinates {(Nrx1, 7.7852) (Nrx2, 4.3762) (Nrx3, 3.3215)};
		
		\end{axis}
		\end{tikzpicture}
		\vspace{-6mm}
		\caption{Fully-analog precoder ($ N_{\mathrm{tx}}^{\mathrm{RF}} = 4 $)}
	\end{subfigure}	
	\vspace{-2mm}
	\caption{Evaluation of e-SINR and latency for different precoders}
	\label{f3}
	\vspace{-5mm}
	\end{minipage}
	\hspace{-5mm}
	\begin{minipage}{0.01\textwidth}
	\end{minipage}
	\begin{minipage}{0.08\textwidth}
	\hspace{1mm}
	\vspace{7mm}
	\centering
	\begin{tikzpicture}[inner sep = 0.3mm]
		\begin{axis}[
		ybar,
		hide axis,
		width = 1.8cm,
		height = 1.8cm,
		xmin = 0,
		xmax = 1,
		ymin = 0,
		ymax = 1,
		legend columns = 1,
		legend style={at={(1,0.5)}, draw=white!15!black, anchor=south west, font=\fontsize{6}{5}\selectfont, text width=1.1cm,text height=4.1mm,text depth=.ex, fill = none},
		]
		
		\addlegendimage{fill = bcolor1, mark=none, line width=0.5pt},
		\addlegendentry{$K_T = 4$},
		\addlegendimage{fill = bcolor2, mark=none, line width=0.5pt},
		\addlegendentry{$K_T = 8$},
		\addlegendimage{fill = bcolor3, mark=none, line width=0.5pt},
		\addlegendentry{$K_T = 16$},
		\addlegendimage{fill = bcolor4, mark=none, line width=0.5pt},
		\addlegendentry{$K_T = 32$},
		\addlegendimage{fill = bcolor5, mark=none, line width=0.5pt},
		\addlegendentry{$K_T = 64$},
		
		\end{axis}
		\end{tikzpicture}
	
	\end{minipage}
	\vspace{-1mm}
\end{figure*}
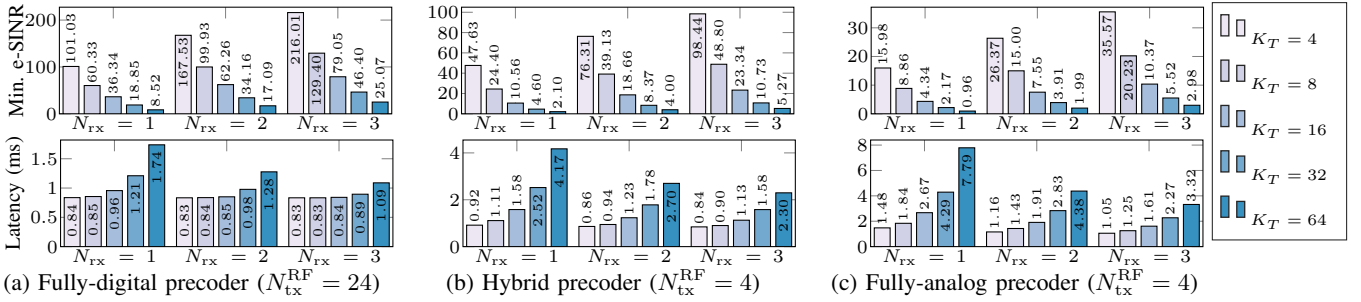

\begin{table}[htb!]
	\scriptsize
	\caption{Simulations configuration}
	\vspace{-2mm}
	\centering
	\begin{tabular}{|l|l|}
		\hline
		{\bf \centering Parameter}				& {\bf Notation \& Value }	\\ 
		\hline
		\hline
		Number of antennas at the transmitter   & $ N_\mathrm{tx} = 24 $ (with $ L_\mathrm{tx} = 6 $) \\ 
		\hline
		Number of antennas at the receiver      & $ N_\mathrm{rx} = \left\lbrace 1, 2, 3 \right\rbrace $ \\ 
		\hline
		Number of RF chains	at the transmitter  & $ N^\mathrm{RF}_\mathrm{tx} = 4 $ \\ 
		\hline
		Number of shifts at the transmitter	    & $ D_F = 16 $ \\ 
		\hline
		Number of phase shifts at the receivers	& $ D_W= 4 $ \\ 
		\hline
		Maximum transmit power		& $ P^{\mathrm{max}}_\mathrm{tx} = 20 $ dBm ($ 100 $ mW) \\ 
		\hline
		Maximum receive power		& $ P^{\mathrm{max}}_\mathrm{rx} = 0 $ dBm ($ 1 $ mW) \\ 
		\hline
		Noise power					& $ \sigma^2 = 10 $ dBm \\ 
		\hline
		Number of multicast groups  & $ G_T = 4 $ \\ 
		\hline
		Number of receivers		    & $ K_T = \left\lbrace 4, 8, 16, 32, 64 \right\rbrace $ \\ 
		\hline
		Bit-stream length		    & $ B_{1} = B_{2} = B_{3} = B_{4} = 4$ Mbits \\ 
		\hline
		Number of bisection procedures & $ N_{\mathrm{bis}_1} = N_{\mathrm{bis}_2} = N_{\mathrm{bis}_3} = 10 $ \\ 
		\hline
		\multirow{3}{*}{Number of randomizations} & $ N_{\mathrm{rand}_1} = 5 \cdot N_\mathrm{tx} $ \\ 
												  & $ N_{\mathrm{rand}_2} = 100 \cdot | \mathcal{G}_t | $ \\										    
												  & $ N_{\mathrm{rand}_3} = 20 \cdot N_\mathrm{rx} $ \\ 
		\hline
		Number of sequential iterations & $ N_\mathrm{iter} = 3 $ \\ 
		\hline
	\end{tabular}
	\label{tab:param}
	\vspace{-3mm}
\end{table}

\subsubsection{Performance of Hybrid Precoder without Scheduling}
We compare the performance of the proposed hybrid precoder design against a fully-digital implementation, which can be obtained as a particular case of our formulation when $ N^{\mathrm{RF}}_\mathrm{tx} = N_\mathrm{tx} $, $ L_\mathrm{tx} = 1 $ and $ \mathbf{F} = \mathbf{I} $. Similarly, we also consider a fully-analog precoder that can be obtained when $ \mathbf{M} = \mathbf{I} $. In this scenario, we consider a single window $ t' $ with a defined scheduling $ \mathcal{V}_{t'} $, where two multicast groups are concurrently served. The performance of this scenario is shown in Fig. \ref{f3} and is essentially related to \emph{Proposition 2}. We observe that, for a given number of receivers $ K_T $, the minimum e-SINR improves with an increasing number of receive antennas $ N_\mathrm{rx} $, because the beam-steering capability of each receiver is augmented. Further, for a given $ N_\mathrm{rx} $, as the number of receivers $ K_T $ increases, the e-SINR reduces since the total power is distributed accordingly. In terms of latency, (calculated with (\ref{e4})), the fully-digital precoder outperforms the hybrid and fully-analog precoders as it is endowed with a larger amount of RF chains that allows enhanced interference reduction, thus promoting higher data rates. However, due to excessive power consumption and hardware costs of fully-digital precoders in mmWave frequencies, the hybrid precoder is a promising candidate with low power consumption and high performance. For instance, if $ P_\mathrm{RF} = 250 $ mW \cite{b17} and $ P_\mathrm{PS} = 30 $ mW \cite{b18} represent the power consumed by a single RF chain and a single 4-bit-resolution phase shifter ($ D_F = 16 $), the instantaneous power consumed by the hybrid and digital precoders are $ P_\mathrm{hyb} = P^{\mathrm{max}}_\mathrm{tx} + N^{\mathrm{RF}}_\mathrm{tx} P_\mathrm{RF} + N_\mathrm{tx} P_\mathrm{PS} = 1.82 $ W and $ P_\mathrm{dig} = P^{\mathrm{max}}_\mathrm{tx} + N^{\mathrm{RF}}_\mathrm{tx} P_\mathrm{RF} = 6.10 $ W, respectively. This reveals an improvement of $ 235\% $ on energy consumption and $ 44\%-232\% $ on energy efficiency (results excluded due to space limitation).

\subsubsection{Performance of Hybrid Precoder with Scheduling}
\begin{figure*}[!t]
	\vspace{-1.5mm}
	\begin{subfigure}[t]{0.33\textwidth}
		\begin{tikzpicture}
			[
			every axis/.style={ 
			ybar stacked,
			ymin = 0,
			ymax = 5.1,
			width = 7.0cm,
			height = 3.85cm,
			bar width = 4.5pt,
			tick align = inside,
			ytick style = {draw = none},
			xtick style = {draw = none},
			x label style={align=center, font=\footnotesize,},
			ylabel = {Latency (ms)},
			y label style={at={(-0.04,0.5)}, font=\footnotesize,},
			symbolic x coords = {K4, K8, K16},
			xticklabels = {$ K_T = 16 $, $ K_T = 32 $, $ K_T = 64 $},
			x tick label style = {text width = 1.7cm, align = center, font = \fontsize{7}{8}\selectfont},
			y tick label style = {text width = 0.3cm, align = right, font = \fontsize{7}{8}\selectfont},
			xtick = data,
			enlarge y limits = {value = 0.3, upper},
			enlarge x limits = 0.24,
			nodes near coords,
			nodes near coords align = {vertical},
            every node near coord/.append style={xshift=\xdelta,yshift=\ydelta-10pt, color = black, rotate = 90, anchor = west, font = \fontsize{2}{2}\selectfont},
			legend columns = -1,
			legend pos = north east,
			legend style={at={(-0.13,1.02)},anchor=south west, font=\fontsize{5}{4}\selectfont, text height=0.02cm, text depth = .ex, fill = none, /tikz/every even column/.append style={column sep=0.046cm}}},
			show sum on top/.style={ 
            /pgfplots/scatter/@post marker code/.append code={%
                \node[
                at={(normalized axis cs:%
                    \pgfkeysvalueof{/data point/x},%
                    \pgfkeysvalueof{/data point/y})%
                },
                xshift=\xdelta, yshift=\ydelta, anchor=west, color = black, rotate = 90, font = \fontsize{2}{2}\selectfont,
                ]
                {\pgfmathprintnumber{\pgfkeysvalueof{/data point/y}}};
            	},  
            },
			]		
		
			\begin{axis}[bar shift = -22.0pt, hide axis]
				\addlegendimage{gcolor7!80, fill = dcolor1, mark=none, line width=0.5pt},		
				\addlegendentry{\texttt{SING}},
				
				\addlegendimage{gcolor7!80, fill = dcolor2, mark=none, line width=0.5pt},
				\addlegendentry{\texttt{RAND}},
				
				\addlegendimage{gcolor7!80, fill = dcolor3, mark=none, line width=0.5pt},
				\addlegendentry{\texttt{XHAUS}},
			
				\addlegendimage{gcolor7!80, fill = dcolor5, mark=none, line width=0.5pt},
				\addlegendentry{\texttt{HYDRAWAVE} $| \lambda = 0.01 $},
			
					
				\addlegendimage{gcolor7!80, fill = dcolor7, mark=none, line width=0.5pt},
				\addlegendentry{\texttt{HYDRAWAVE} $| \lambda = 0.05 $},		
			
				\addlegendimage{gcolor7!80, fill = dcolor8, mark=none, line width=0.5pt},
				\addlegendentry{\texttt{HYDRAWAVE} $| \lambda = 0.1 $},
			
				\addlegendimage{gcolor7!80, fill = dcolor9, mark=none, line width=0.5pt},
				\addlegendentry{\texttt{HYDRAWAVE} $| \lambda = 0.2 $},
			
				\addlegendimage{gcolor7!80, fill = dcolor10, mark=none, line width=0.5pt},
				\addlegendentry{\texttt{HYDRAWAVE} $| \lambda = 0.4 $},
	
				\addlegendimage{gcolor7!80, fill = dcolor11, mark=none, line width=0.5pt},
				\addlegendentry{\texttt{HYDRAWAVE} $| \lambda_\mathrm{opt} $},
				
				\addlegendimage{gcolor7!60, pattern = crosshatch, pattern color = gcolor7!80},
				\addlegendentry{Switching delay},

				\addplot+[visualization depends on={value \thisrow{Cx} \as \xdelta},
              		visualization depends on={value \thisrow{Cy} \as \ydelta},
					every node near coord/.append style={color=white},
					gcolor7!80, fill = dcolor1]
     			table [x = {x}, y = {C}] 
     			{
	      			x 	C       	Cx	Cy 
	      			K4 	3.3352   	0   5  
	      			K8 	3.3487   	0   5.5  
	      			K16 3.5837   	0   6  
    			};
    			\addplot+[visualization depends on={value \thisrow{Dx} \as \xdelta},
              		visualization depends on={value \thisrow{Dy} \as \ydelta},
					every node near coord/.append style={opacity=0},
              		show sum on top,
					gcolor7!60, pattern = crosshatch, pattern color = dcolor1] 
     			table [x = {x}, y = {D}] 
     			{
	      			x 	D		Dx  	Dy
	      			K4 	1.5    	-22.0   -2 
	      			K8 	1.5    	-22.0   -2 
	      			K16 1.5    	-22.0   -2
    			};
				
			\end{axis}	
			
			\begin{axis}[bar shift = -16.5pt, hide axis]				
				\addplot+[visualization depends on={value \thisrow{Cx} \as \xdelta},
              		visualization depends on={value \thisrow{Cy} \as \ydelta},
					gcolor7, fill = dcolor2]
     			table [x = {x}, y = {C}] 
     			{
	      			x 	C        Cx  Cy 
	      			K4 	2.2167   0   2 
	      			K8 	2.5084   0   2.5  
	      			K16 3.5588   0   6  
    			};
    			\addplot+[visualization depends on={value \thisrow{Dx} \as \xdelta},
              		visualization depends on={value \thisrow{Dy} \as \ydelta},
					every node near coord/.append style={opacity=0},
              		show sum on top,
					gcolor7!60, pattern = crosshatch, pattern color = dcolor2] 
     			table [x = {x}, y = {D}] 
     			{
	      			x 	D	 	Dx  	Dy
	      			K4  0.7450	-16.5   -2 
	      			K8  0.7050  -16.5   -2 
	      			K16 0.6900  -16.5   -2
    			};	
			\end{axis}        
			
			\begin{axis}[bar shift = -11.0pt, hide axis]
				\addplot+[visualization depends on={value \thisrow{Cx} \as \xdelta},
              		visualization depends on={value \thisrow{Cy} \as \ydelta},
					gcolor7!80, fill = dcolor3]
     			table [x = {x}, y = {C}] 
     			{
      			x 	C 		Cx  Cy 
      			K4 	1.3216  0  	0  
      			K8 	1.9241  0   1  
      			K16 2.5541 	0   4  
    			};
    			\addplot+[visualization depends on={value \thisrow{Dx} \as \xdelta},
              		visualization depends on={value \thisrow{Dy} \as \ydelta},
					every node near coord/.append style={opacity=0},
              		show sum on top,
					gcolor7!60, pattern = crosshatch, pattern color = dcolor3] 
     			table [x = {x}, y = {D}] 
     			{
	      			x 	D    	Dx  	Dy
	      			K4 	0.0130  -11.0	-2 
	      			K8 	0.2099  -11.0	-2 
	      			K16 0.4949 	-11.0	-2
    			};	
			\end{axis}        
	
			\begin{axis}[bar shift = -5.5pt, hide axis]
				\addplot+[visualization depends on={value \thisrow{Cx} \as \xdelta},
              		visualization depends on={value \thisrow{Cy} \as \ydelta},
					gcolor7!80, fill = dcolor5]
     			table [x = {x}, y = {C}] 
     			{
	      			x 	C      	Cx  Cy 
	      			K4 	3.1382  0   6  
	      			K8 	3.1683  0   6  
	      			K16 3.4699 	0   7  
    			};
    			\addplot+[visualization depends on={value \thisrow{Dx} \as \xdelta},
              		visualization depends on={value \thisrow{Dy} \as \ydelta},
					every node near coord/.append style={opacity=0},
              		show sum on top,
					gcolor7!60, pattern = crosshatch, pattern color = dcolor5] 
     			table [x = {x}, y = {D}] 
     			{
	      			x 	D   	Dx  	Dy
	      			K4 	1.3800  -5.5	-2 
	      			K8 	1.3750  -5.5	-2 
	      			K16 1.3850 	-5.5	-2
    			};	
			\end{axis}        
			
 
			\begin{axis}[bar shift = 0.0pt, hide axis]
				\addplot+[visualization depends on={value \thisrow{Cx} \as \xdelta},
              		visualization depends on={value \thisrow{Cy} \as \ydelta},
					gcolor7!80, fill = dcolor7]
     			table [x = {x}, y = {C}] 
     			{
	      			x 	C      	Cx  Cy 
	      			K4 	2.2945  0  	2  
	      			K8 	2.4499  0  	3  
	      			K16 3.0175 	0  	5  
    			};
    			\addplot+[visualization depends on={value \thisrow{Dx} \as \xdelta},
              		visualization depends on={value \thisrow{Dy} \as \ydelta},
					every node near coord/.append style={opacity=0},
              		show sum on top,
					gcolor7!60, pattern = crosshatch, pattern color = dcolor7] 
     			table [x = {x}, y = {D}] 
     			{
	      			x 	D   	Dx  Dy
	      			K4 	0.8650 	0.0	-2 
	      			K8 	0.8550  0.0	-2 
	      			K16 0.8750 	0.0	-2
    			};	
			\end{axis}
			
			\begin{axis}[bar shift = 5.5pt, hide axis]
				\addplot+[visualization depends on={value \thisrow{Cx} \as \xdelta},
              		visualization depends on={value \thisrow{Cy} \as \ydelta},
					gcolor7!80, fill = dcolor8]
     			table [x = {x}, y = {C}]  
     			{
	      			x 	C      	Cx  Cy 
	      			K4 	1.8670  0   0  
	      			K8 	2.1433  0   1  
	      			K16 2.8906 	0   5  
    			};
    			\addplot+[visualization depends on={value \thisrow{Dx} \as \xdelta},
              		visualization depends on={value \thisrow{Dy} \as \ydelta},
					every node near coord/.append style={opacity=0},
              		show sum on top,
					gcolor7!60, pattern = crosshatch, pattern color = dcolor8] 
     			table [x = {x}, y = {D}] 
     			{
	      			x 	D   	Dx  Dy
	      			K4 	0.5950  5.5	-2 
	      			K8 	0.6000  5.5	-2 
	      			K16 0.6450 	5.5	-2
    			};	
			\end{axis}       
					
			\begin{axis}[bar shift = 11.0pt, hide axis]
				\addplot+[visualization depends on={value \thisrow{Cx} \as \xdelta},
              		visualization depends on={value \thisrow{Cy} \as \ydelta},
					every node near coord/.append style={color=white},
					gcolor7!80, fill = dcolor9] 
     			table [x = {x}, y = {C}] 
     			{
	      			x 	C      	Cx  Cy 
	      			K4 	1.7234  0   0  
	      			K8 	2.0202  0   0  
	      			K16 2.8355 	0   4  
    			};
    			\addplot+[visualization depends on={value \thisrow{Dx} \as \xdelta},
              		visualization depends on={value \thisrow{Dy} \as \ydelta},
					every node near coord/.append style={opacity=0},
              		show sum on top,
					gcolor7!60, pattern = crosshatch, pattern color = dcolor9] 
     			table [x = {x}, y = {D}] 
     			{
	      			x 	D   	Dx		Dy
	      			K4 	0.5000  11.0	-2 
	      			K8 	0.4950  11.0	-2 
	      			K16 0.5050 	11.0	-2
    			};
			\end{axis}
			
			\begin{axis}[bar shift = 16.5pt, hide axis]
				\addplot+[visualization depends on={value \thisrow{Cx} \as \xdelta},
              		visualization depends on={value \thisrow{Cy} \as \ydelta},
					every node near coord/.append style={color=white},
					gcolor7!80, fill = dcolor10]
     			table [x = {x}, y = {C}] 
     			{
	      			x 	C      	Cx  Cy 
	      			K4 	1.6612  0   0  
	      			K8 	2.0380  0   0  
	      			K16 3.0265 	0   4  
    			};
    			\addplot+[visualization depends on={value \thisrow{Dx} \as \xdelta},
              		visualization depends on={value \thisrow{Dy} \as \ydelta},
					every node near coord/.append style={opacity=0},
              		show sum on top,
					gcolor7!60, pattern = crosshatch, pattern color = dcolor10] 
     			table [x = {x}, y = {D}] 
     			{
	      			x 	D   	Dx  	Dy
	      			K4 	0.4250  16.5	-2 
	      			K8 	0.4450  16.5	-2 
	      			K16 0.4500 	16.5	-2
    			};
			\end{axis}        	
			
			\begin{axis}[bar shift = 22.0pt]
				\addplot+[visualization depends on={value \thisrow{Cx} \as \xdelta},
              		visualization depends on={value \thisrow{Cy} \as \ydelta},
					every node near coord/.append style={color=white},
					gcolor7!80, fill = dcolor11]
     			table [x = {x}, y = {C}] 
     			{
	      			x 	C   	Cx Cy 
	      			K4 	1.3236  0  0  
	      			K8 	2.0202  0  0  
	      			K16 2.8345 	0  4  
    			};
    			\addplot+[visualization depends on={value \thisrow{Dx} \as \xdelta},
              		visualization depends on={value \thisrow{Dy} \as \ydelta},
					every node near coord/.append style={opacity=0},
              		show sum on top,
					gcolor7!60, pattern = crosshatch, pattern color = dcolor11] 
     			table [x = {x}, y = {D}] 
     			{
	      			x 	D   	Dx  	Dy
	      			K4 	0.0100  22.0	-2 
	      			K8 	0.1624  22.0	-2 
	      			K16 0.5000 	22.0	-2
    			};
			\end{axis}
		
		\end{tikzpicture}
		\vspace{-6mm}
		\centering
		\subcaption{Fully-digital precoder}
	\end{subfigure}%
	\begin{subfigure}[t]{0.33\textwidth}
		\begin{tikzpicture}
			[
			every axis/.style={ 
			ybar stacked,
			ymin = 0,
			ymax = 8.7,
			width = 7.0cm,
			height = 3.85cm,
			bar width = 4.5pt,
			tick align = inside,
			ytick style = {draw = none},
			xtick style = {draw = none},
			x label style={align=center, font=\footnotesize,},
			symbolic x coords = {K4, K8, K16},
			xticklabels = {$ K_T = 16 $, $ K_T = 32 $, $ K_T = 64 $},
			x tick label style = {text width = 1.7cm, align = center, font = \fontsize{7}{8}\selectfont},
			y tick label style = {text width = 0.3cm, align = right, font = \fontsize{7}{8}\selectfont},
			xtick = data,
			enlarge y limits = {value = 0.3, upper},
			enlarge x limits = 0.24,
			nodes near coords,
			nodes near coords align = {vertical},
            every node near coord/.append style={xshift=\xdelta,yshift=\ydelta-8pt, color = black, rotate = 90, anchor = west, font = \fontsize{2}{2}\selectfont},
			legend columns = -1,
			legend pos = north east,
			legend style={at={(0.00,1.08)},anchor=south west, font=\fontsize{5}{4}\selectfont, text height=0.02cm, text depth = .ex, fill = none, /tikz/every even column/.append style={column sep=0.0525cm}}},
			show sum on top/.style={ 
            /pgfplots/scatter/@post marker code/.append code={%
                \node[
                at={(normalized axis cs:%
                    \pgfkeysvalueof{/data point/x},%
                    \pgfkeysvalueof{/data point/y})%
                },
                xshift=\xdelta, yshift=\ydelta, anchor=west, color = black, rotate = 90, font = \fontsize{2}{2}\selectfont,
                ]
                {\pgfmathprintnumber{\pgfkeysvalueof{/data point/y}}};
            	},  
            },
			]
			
			\begin{axis}[bar shift = -22.0pt, hide axis]
				\addplot+[visualization depends on={value \thisrow{Cx} \as \xdelta},
              		visualization depends on={value \thisrow{Cy} \as \ydelta},
					every node near coord/.append style={color=white},
					gcolor7!80, fill = dcolor1] 
     			table [x = {x}, y = {C}] 
     			{
	      			x 	C      	Cx	Cy 
	      			K4 	3.3384	0  	0  
	      			K8 	3.4198	0  	0  
	      			K16 3.9737 	0  	0  
    			};
    			\addplot+[visualization depends on={value \thisrow{Dx} \as \xdelta},
              		visualization depends on={value \thisrow{Dy} \as \ydelta},
					every node near coord/.append style={opacity=0},
              		show sum on top,
					gcolor7!60, pattern = crosshatch, pattern color = dcolor1] 
					table [x = {x}, y = {D}] 
					{
		      			x 	D 		Dx  	Dy
		      			K4 	1.5  	-22.0	-2 
		      			K8 	1.5  	-22.0	-2 
		      			K16 1.5 	-22.0	-2
	    			};
			\end{axis}
			
			\begin{axis}[bar shift = -16.5pt, hide axis]
				\addplot+[visualization depends on={value \thisrow{Cx} \as \xdelta},
              		visualization depends on={value \thisrow{Cy} \as \ydelta},
					gcolor7!80, fill = dcolor2] 
     			table [x = {x}, y = {C}] 
     			{
	      			x 	C   	Cx  Cy
	      			K4 	3.1692  0   0 
	      			K8 	4.8318  0   0 
	      			K16 8.0442 	0   0  
    			};
    			\addplot+[visualization depends on={value \thisrow{Dx} \as \xdelta},
              		visualization depends on={value \thisrow{Dy} \as \ydelta},
					every node near coord/.append style={opacity=0},
              		show sum on top,
					gcolor7!60, pattern = crosshatch, pattern color = dcolor2] 
     			table [x = {x}, y = {D}]  
     			{
	      			x 	D   	Dx  	Dy
	      			K4 	0.7450  -16.5	-2 
	      			K8 	0.7050  -16.5	-2 
	      			K16 0.6900 	-16.5	-2
    			};
			\end{axis} 
			    
			\begin{axis}[bar shift = -11.0pt, hide axis]
				\addplot+[visualization depends on={value \thisrow{Cx} \as \xdelta},
              		visualization depends on={value \thisrow{Cy} \as \ydelta},
					gcolor7!80, fill = dcolor3]
     			table [x = {x}, y = {C}] 
     			{
	      			x 	C      	Cx  Cy 
	      			K4 	2.3701  0  	-2  
	      			K8 	3.1827  0  	0  
	      			K16 3.9117 	0  	0  
    			};
    			\addplot+[visualization depends on={value \thisrow{Dx} \as \xdelta},
              		visualization depends on={value \thisrow{Dy} \as \ydelta},
					every node near coord/.append style={opacity=0},
              		show sum on top,
					gcolor7!60, pattern = crosshatch, pattern color = dcolor3] 
     			table [x = {x}, y = {D}] 
     			{
	      			x 	D   	Dx  	Dy
	      			K4 	0.5023  -11.0	-2 
	      			K8 	0.7250  -11.0	-2 
	      			K16 1.1959 	-11.0	-2
    			};
			\end{axis}        
					
			\begin{axis}[bar shift = -5.5pt, hide axis]
			\addplot+[visualization depends on={value \thisrow{Cx} \as \xdelta},
              		visualization depends on={value \thisrow{Cy} \as \ydelta},
					gcolor7!80, fill = dcolor5]
     			table [x = {x}, y = {C}] 
     			{
	      			x 	C      	Cx  Cy 
	      			K4 	3.2291 	0  	0  
	      			K8 	3.4243 	0  	0  
	      			K16 4.1184 	0  	0  
    			};
    			\addplot+[visualization depends on={value \thisrow{Dx} \as \xdelta},
              		visualization depends on={value \thisrow{Dy} \as \ydelta},
					every node near coord/.append style={opacity=0},
              		show sum on top,
					gcolor7!60, pattern = crosshatch, pattern color = dcolor5] 
     			table [x = {x}, y = {D}] 
     			{
	      			x 	D   	Dx  	Dy
	      			K4 	1.3800  -5.5	-2 
	      			K8 	1.3750  -5.5	-2 
	      			K16 1.3850 	-5.5	-2
    			};
			\end{axis}	       
	
				
			\begin{axis}[bar shift = 0.0pt, hide axis]
				\addplot+[visualization depends on={value \thisrow{Cx} \as \xdelta},
              		visualization depends on={value \thisrow{Cy} \as \ydelta},
					gcolor7!80, fill = dcolor7]
     			table [x = {x}, y = {C}] 
     			{
	      			x 	C      	Cx  Cy 
	      			K4 	2.7543  0  	0  
	      			K8 	3.5078  0  	0  
	      			K16 5.0191 	0  	0  
    			};
    			\addplot+[visualization depends on={value \thisrow{Dx} \as \xdelta},
              		visualization depends on={value \thisrow{Dy} \as \ydelta},
					every node near coord/.append style={opacity=0},
              		show sum on top,
					gcolor7!60, pattern = crosshatch, pattern color = dcolor7] 
     			table [x = {x}, y = {D}] 
     			{
	      			x 	D   	Dx 	Dy
	      			K4 	0.8650  0.0	-2 
	      			K8 	0.8550  0.0	-2 
	      			K16 0.8750 	0.0	-2
    			};
			\end{axis}	        
			
			\begin{axis}[bar shift = 5.5pt, hide axis]
				\addplot+[visualization depends on={value \thisrow{Cx} \as \xdelta},
              		visualization depends on={value \thisrow{Cy} \as \ydelta},
					gcolor7!80, fill = dcolor8]
     			table [x = {x}, y = {C}] 
     			{
	      			x 	C      	Cx  Cy 
	      			K4 	2.5876  0  	0  
	      			K8 	3.6754  0  	0  
	      			K16 5.6329 	0  	0  
    			};
    			\addplot+[visualization depends on={value \thisrow{Dx} \as \xdelta},
              		visualization depends on={value \thisrow{Dy} \as \ydelta},
					every node near coord/.append style={opacity=0},
              		show sum on top,
					gcolor7!60, pattern = crosshatch, pattern color = dcolor8] 
     			table [x = {x}, y = {D}] 
     			{
	      			x 	D   	Dx		Dy
	      			K4 	0.5950  5.5		-2 
	      			K8 	0.6000  5.5		-2 
	      			K16 0.6450 	5.5		-2
    			};
			\end{axis}	        	
			
			\begin{axis}[bar shift = 11.0pt, hide axis]
				\addplot+[visualization depends on={value \thisrow{Cx} \as \xdelta},
              		visualization depends on={value \thisrow{Cy} \as \ydelta},
					every node near coord/.append style={color=white},
					gcolor7!80, fill = dcolor9]
     			table [x = {x}, y = {C}] 
     			{
	      			x 	C   	Cx  Cy
	      			K4 	2.5311  0   0 
	      			K8 	3.8623  0   0 
	      			K16 6.2670 	0   0 
    			};
    			\addplot+[visualization depends on={value \thisrow{Dx} \as \xdelta},
              		visualization depends on={value \thisrow{Dy} \as \ydelta},
					every node near coord/.append style={opacity=0},
              		show sum on top,
					gcolor7!60, pattern = crosshatch, pattern color = dcolor9] 
     			table [x = {x}, y = {D}] 
     			{
	      			x 	D   	Dx  	Dy
	      			K4 	0.5000  11.0	-2
	      			K8 	0.4950  11.0	-2 
	      			K16 0.5050 	11.0	-2
    			};
			\end{axis}
			
			\begin{axis}[bar shift = 16.5pt, hide axis]
				\addplot+[visualization depends on={value \thisrow{Cx} \as \xdelta},
              		visualization depends on={value \thisrow{Cy} \as \ydelta},
					every node near coord/.append style={color=white},
					gcolor7!80, fill = dcolor10]
     			table [x = {x}, y = {C}] 
     			{
	      			x 	C   	Cx  Cy
	      			K4 	2.9734  0   0 
	      			K8 	4.6320  0   0 
	      			K16 7.7664 	0   0 
    			};
    			\addplot+[visualization depends on={value \thisrow{Dx} \as \xdelta},
              		visualization depends on={value \thisrow{Dy} \as \ydelta},
					every node near coord/.append style={opacity=0},
              		show sum on top,
					gcolor7!60, pattern = crosshatch, pattern color = dcolor10] 
     			table [x = {x}, y = {D}] 
     			{
	      			x 	D   	Dx  	Dy
	      			K4 	0.4250  16.5	-2 
	      			K8 	0.4450  16.5	-2 
	      			K16 0.4500 	16.5	-2
    			};
			\end{axis}        
	
			\begin{axis}[bar shift = 22.0pt]
				\addplot+[visualization depends on={value \thisrow{Cx} \as \xdelta},
              		visualization depends on={value \thisrow{Cy} \as \ydelta},
					every node near coord/.append style={color=white},
					gcolor7!80, fill = dcolor11] 
     			table [x = {x}, y = {C}] 
     			{
	      			x 	C   	Cx  Cy
	      			K4 	2.5311  0   0 
	      			K8 	3.4087  0   0 
	      			K16 3.9737 	0   0 
    			};
    			\addplot+[visualization depends on={value \thisrow{Dx} \as \xdelta},
              		visualization depends on={value \thisrow{Dy} \as \ydelta},
					every node near coord/.append style={opacity=0},
              		show sum on top,
					gcolor7!60, pattern = crosshatch, pattern color = dcolor11] 
     			table [x = {x}, y = {D}] 
     			{
	      			x 	D   	Dx  	Dy
	      			K4 	0.5000  22.0	-2 
	      			K8 	0.8667  22.0	-2 
	      			K16 1.4962 	22.0	-2
    			};
			\end{axis}		
		
		\end{tikzpicture}
		\vspace{-6mm}
		\subcaption{Hybrid precoder}
	\end{subfigure}%
	\begin{subfigure}[t]{0.33\textwidth}
		\begin{tikzpicture}
			[
			every axis/.style={ 
			ybar stacked,
			ymin = 0,
			ymax = 14.0,
			width = 7.0cm,
			height = 3.85cm,
			bar width = 4.5pt,
			tick align = inside,
			ytick style = {draw = none},
			xtick style = {draw = none},
			x label style={align=center, font=\footnotesize,},
			nodes near coords,
			nodes near coords align = {vertical},
            every node near coord/.append style={xshift=\xdelta,yshift=\ydelta-9pt, color = black, rotate = 90, anchor = west, font = \fontsize{2}{2}\selectfont},
			symbolic x coords = {K4, K8, K16},
			xticklabels = {$ K_T = 16 $, $ K_T = 32 $, $ K_T = 64 $},
			x tick label style = {text width = 1.7cm, align = center, font = \fontsize{7}{8}\selectfont},
			y tick label style = {text width = 0.3cm, align = right, font = \fontsize{7}{8}\selectfont},
			xtick = data,
			enlarge y limits = {value = 0.3, upper},
			enlarge x limits = 0.24,
			legend columns = -1,
			legend pos = north east,
			legend style={at={(0.00,1.08)},anchor=south west, font=\fontsize{5}{4}\selectfont, text height=0.02cm, text depth = .ex, fill = none, /tikz/every even column/.append style={column sep=0.0525cm}}},
			show sum on top/.style={ 
            /pgfplots/scatter/@post marker code/.append code={%
                \node[
                at={(normalized axis cs:%
                    \pgfkeysvalueof{/data point/x},%
                    \pgfkeysvalueof{/data point/y})%
                },
                xshift=\xdelta, yshift=\ydelta, anchor=west, color = black, rotate = 90, font = \fontsize{2}{2}\selectfont,
                ]
                {\pgfmathprintnumber{\pgfkeysvalueof{/data point/y}}};
            	},  
            },
			]	

			\begin{axis}[bar shift = -22.0pt, hide axis]
				\addplot+[visualization depends on={value \thisrow{Cx} \as \xdelta},
              		visualization depends on={value \thisrow{Cy} \as \ydelta},
					every node near coord/.append style={color=white},
					gcolor7!80, fill = dcolor1]
     			table [x = {x}, y = {C}] 
     			{
	      			x 	C      	Cx   Cy 
	      			K4 	4.2027	0    0  
	      			K8 	5.1846	0    0  
	      			K16 6.4728 	0    0  
    			};
    			\addplot+[visualization depends on={value \thisrow{Dx} \as \xdelta},
              		visualization depends on={value \thisrow{Dy} \as \ydelta},
					every node near coord/.append style={opacity=0},
              		show sum on top,
					gcolor7!60, pattern = crosshatch, pattern color = dcolor1] 
     			table [x = {x}, y = {D}] 
     			{
	      			x 	D    Dx  	Dy
	      			K4 	1.5	 -22	-2 
	      			K8 	1.5  -22	-2 
	      			K16 1.5  -22	-2
    			};
			\end{axis}
			
			\begin{axis}[bar shift = -16.5pt, hide axis]
				\addplot+[visualization depends on={value \thisrow{Cx} \as \xdelta},
              		visualization depends on={value \thisrow{Cy} \as \ydelta},
					gcolor7!80, fill = dcolor2] 
     			table [x = {x}, y = {C}] 
     			{
	      			x 	C   	Cx  Cy
	      			K4 	4.7238 	0   0 
	      			K8 	7.4616  0   0 
	      			K16 12.4183 0   0 
    			};
    			\addplot+[visualization depends on={value \thisrow{Dx} \as \xdelta},
              		visualization depends on={value \thisrow{Dy} \as \ydelta},
					every node near coord/.append style={opacity=0},
              		show sum on top,
					gcolor7!60, pattern = crosshatch, pattern color = dcolor2] 
     			table [x = {x}, y = {D}] 
     			{
	      			x 	D   	Dx		Dy
	      			K4 	0.745  	-16.5	-2 
	      			K8 	0.705  	-16.5	-2 
	      			K16 0.690 	-16.5	-2
    			};
			\end{axis}
			
			\begin{axis}[bar shift = -11.0pt, hide axis]
				\addplot+[visualization depends on={value \thisrow{Cx} \as \xdelta},
              		visualization depends on={value \thisrow{Cy} \as \ydelta},
					gcolor7!80, fill = dcolor3] 
     			table [x = {x}, y = {C}] 
     			{
	      			x 	C   	Cx  Cy 
	      			K4 	3.4959 	0  	0  
	      			K8 	4.8673 	0  	0  
	      			K16 6.2551 	0  	0  
    			};
    			\addplot+[visualization depends on={value \thisrow{Dx} \as \xdelta},
              		visualization depends on={value \thisrow{Dy} \as \ydelta},
					every node near coord/.append style={opacity=0},
              		show sum on top,
					gcolor7!60, pattern = crosshatch, pattern color = dcolor3] 
     			table [x = {x}, y = {D}]  
     			{
	      			x 	D   	Dx		Dy
	      			K4 	0.6038  -11.0	-2
	      			K8 	0.8634  -11.0	-2 
	      			K16 1.1952 	-11.0	-2
    			};
			\end{axis}
					
			\begin{axis}[bar shift = -5.5pt, hide axis]
				\addplot+[visualization depends on={value \thisrow{Cx} \as \xdelta},
              		visualization depends on={value \thisrow{Cy} \as \ydelta},
					gcolor7!80, fill = dcolor5] 
     			table [x = {x}, y = {C}]  
     			{
	      			x 	C   	Cx  Cy 
	      			K4 	4.1664  0  0  
	      			K8 	5.2866  0  0  
	      			K16 6.6371 	0  0  
    			};
    			\addplot+[visualization depends on={value \thisrow{Dx} \as \xdelta},
              		visualization depends on={value \thisrow{Dy} \as \ydelta},
					every node near coord/.append style={opacity=0},
              		show sum on top,
					gcolor7!60, pattern = crosshatch, pattern color = dcolor5] 
     			table [x = {x}, y = {D}] 
     			{
	      			x 	D   	Dx  	Dy
	      			K4 	1.3800  -5.5	-2 
	      			K8 	1.3750  -5.5	-2 
	      			K16 1.3850 	-5.5	-2
    			};
			\end{axis}	
	
				
			\begin{axis}[bar shift = 0.0pt, hide axis]
				\addplot+[visualization depends on={value \thisrow{Cx} \as \xdelta},
              		visualization depends on={value \thisrow{Cy} \as \ydelta},
					gcolor7!80, fill = dcolor7] 
     			table [x = {x}, y = {C}] 
     			{
	      			x 	C      	Cx  Cy 
	      			K4 	3.9119  0   0  
	      			K8 	5.4890  0   0  
	      			K16 8.0463 	0   0  
    			};
    			\addplot+[visualization depends on={value \thisrow{Dx} \as \xdelta},
              		visualization depends on={value \thisrow{Dy} \as \ydelta},
					every node near coord/.append style={opacity=0},
              		show sum on top,
					gcolor7!60, pattern = crosshatch, pattern color = dcolor7] 
     			table [x = {x}, y = {D}] 
     			{
	      			x 	D   	Dx 	Dy
	      			K4 	0.8650  0.0	-2 
	      			K8 	0.8550  0.0	-2 
	      			K16 0.8750 	0.0	-2
    			};
			\end{axis}	
			    
			\begin{axis}[bar shift = 5.5pt, hide axis]
				\addplot+[visualization depends on={value \thisrow{Cx} \as \xdelta},
              		visualization depends on={value \thisrow{Cy} \as \ydelta},
					gcolor7!80, fill = dcolor8] 
     			table [x = {x}, y = {C}] 
     			{
	      			x 	C      	Cx  Cy 
	      			K4 	3.9540  0   0  
	      			K8 	5.8668  0   0  
	      			K16 9.2767 	0   0  
    			};
    			\addplot+[visualization depends on={value \thisrow{Dx} \as \xdelta},
              		visualization depends on={value \thisrow{Dy} \as \ydelta},
					every node near coord/.append style={opacity=0},
              		show sum on top,
					gcolor7!60, pattern = crosshatch, pattern color = dcolor8] 
     			table [x = {x}, y = {D}] 
     			{
	      			x 	D   	Dx  Dy
	      			K4 	0.5950  5.5	-2 
	      			K8 	0.6000  5.5	-2 
	      			K16 0.6450 	5.5	-2
    			};
			\end{axis}	
			    
			\begin{axis}[bar shift = 11.0pt, hide axis]
				\addplot+[visualization depends on={value \thisrow{Cx} \as \xdelta},
              		visualization depends on={value \thisrow{Cy} \as \ydelta},
					every node near coord/.append style={color=white},
					gcolor7!80, fill = dcolor9]
     			table [x = {x}, y = {C}] 
     			{
	      			x 	C      	 Cx  Cy 
	      			K4 	3.9870   0   0  
	      			K8 	6.0888   0   0  
	      			K16 10.2409  0   0  
    			};
    			\addplot+[visualization depends on={value \thisrow{Dx} \as \xdelta},
              		visualization depends on={value \thisrow{Dy} \as \ydelta},
					every node near coord/.append style={opacity=0},
              		show sum on top,
					gcolor7!60, pattern = crosshatch, pattern color = dcolor9] 
     			table [x = {x}, y = {D}] 
     			{
	      			x 	D   	Dx  Dy
	      			K4 	0.5000  11.0	-2 
	      			K8 	0.4950  11.0	-2 
	      			K16 0.5050 	11.0	-2
    			};
			\end{axis}	 	
			
			\begin{axis}[bar shift = 16.5pt, hide axis]
				\addplot+[visualization depends on={value \thisrow{Cx} \as \xdelta},
              		visualization depends on={value \thisrow{Cy} \as \ydelta},
					every node near coord/.append style={color=white},
					gcolor7!80, fill = dcolor10]
     			table [x = {x}, y = {C}] 
     			{
	      			x C      	 Cx  Cy 
	      			K4 	4.4096   0   0  
	      			K8 	6.8653   0   0  
	      			K16 12.0911  0   0  
    			};
    			\addplot+[visualization depends on={value \thisrow{Dx} \as \xdelta},
              		visualization depends on={value \thisrow{Dy} \as \ydelta},
					every node near coord/.append style={opacity=0},
              		show sum on top,
					gcolor7!60, pattern = crosshatch, pattern color = dcolor10] 
     			table [x = {x}, y = {D}]  
     			{
	      			x 	D   	Dx  	Dy
	      			K4 	0.4250  16.5	-2 
	      			K8 	0.4450  16.5	-2 
	      			K16 0.4500 	16.5	-2
    			};
			\end{axis}        
	
			\begin{axis}[bar shift = 22.0pt]
				\addplot+[visualization depends on={value \thisrow{Cx} \as \xdelta},
              		visualization depends on={value \thisrow{Cy} \as \ydelta},
					every node near coord/.append style={color=white},
					gcolor7!80, fill = dcolor11]
     			table [x = {x}, y = {C}] 
     			{
	      			x 	C 		Cx  Cy 
	      			K4 	3.9119 	0   0  
	      			K8 	5.1810  0   0  
	      			K16 6.4653 	0   0  
    			};
    			\addplot+[visualization depends on={value \thisrow{Dx} \as \xdelta},
              		visualization depends on={value \thisrow{Dy} \as \ydelta},
					every node near coord/.append style={opacity=0},
              		show sum on top,
					gcolor7!60, pattern = crosshatch, pattern color = dcolor11] 
     			table [x = {x}, y = {D}] 
     			{
	      			x 	D   	Dx  Dy
	      			K4 	0.5751	22.0	-2 
	      			K8 	1.1630  22.0	-2 
	      			K16 1.4950  22.0	-2
    			};
			\end{axis}				
						
			\end{tikzpicture}
			\vspace{-6mm}
			\subcaption{Fully-analog precoder}
		\end{subfigure}
	\vspace{-2mm}
	\caption{Latency performance of different scheduling schemes with fully-digital, hybrid, and fully-analog precoders.}
	\label{f4}
	\vspace{-2mm}
\end{figure*}
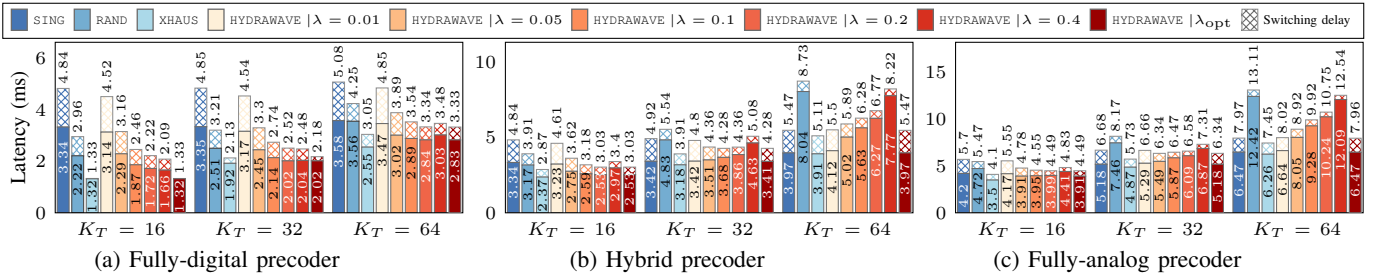

\fref{f4} illustrates the performance of \texttt{\small{HYDRAWAVE}}, \texttt{\small{SING}}, \texttt{\small{RAND}}, and \texttt{\small{XHAUS}}. In our scheduling formulation in (\ref{e21}), $ \lambda $ controls the maximum tolerable aggregate IGC. Thus, in the fully-digital precoder case, a larger $ \lambda $ can be supported due to the versatility of the precoder to manage interference. On the other hand, due to the limited amount of RF chains, the values of $ \lambda $ used with the hybrid and analog precoders need to be comparatively smaller. We observe that, in terms of latency, the fully-digital precoder outperforms the hybrid and fully-analog implementations. Further, for any precoder type with a small number of receivers (e.g., $ K_T = 16 $), \texttt{\small{SING}} produces the greatest latency while \texttt{\small{XHAUS}} is optimal (according to \emph{Proposition 3}). On the other hand, \texttt{\small{RAND}} exhibits an intermediate performance between \texttt{\small{SING}} and \texttt{XHAUS}. As the number of receivers increases (e.g., $ K_T = 64 $), the performance gap between \texttt{\small{SING}} and \texttt{\small{XHAUS}} reduces considerably because interference becomes more complex to manage, thus yielding \texttt{\small{SING}} optimal in some realizations. On the contrary, \texttt{\small{RAND}} deteriorates since the impact of scheduling becomes more relevant in the presence of higher interference. When considering a switching delay of $ \delta_\mathrm{SW} = 0.5 $ ms between consecutive windows \cite{b38}, similar behavior can be observed except for \texttt{\small{SING}} being heavily penalized due to the incapability of spatial multiplexing. Furthermore, we include the performance of \texttt{\small{HYDRAWAVE}} with different values of $ \lambda $. In Fig. \ref{f4}, $ \lambda_{\mathrm{opt}} $ is obtained upon evaluating several $ \lambda $ and bisecting the search space until negligible variation is observed. For the fully-digital precoder, \texttt{\small{HYDRAWAVE}} attains near-optimality when $ K_T = 16 $. Also, its performance remains within $ 11\% $ of the optimal value for $ K_T = \left\lbrace 32, 64 \right\rbrace $ when $ \delta_\mathrm{SW} = 0.0 $, and within $ 9\% $ when $ \delta_\mathrm{SW} = 0.5 $ ms. In the hybrid and fully-analog cases, the performance of \texttt{\small{HYDRAWAVE}} remains in the range $ 1.5 - 9.5 \% $ and $ 3.4 - 11.7 \%$ of the optimal \texttt{\small{XHAUS}}, respectively. In the hybrid precoder case, \texttt{\small{HYDRAWAVE}} exhibits gains up $ 32 \% $ higher than \texttt{\small{SING}} and up to $ 102 \% $ compared to \texttt{\small{RAND}} when $ \delta_\mathrm{SW} = 0 $. When $ \delta_\mathrm{SW} = 0.5 $, the gains are up to $ 60 \% $ and $ 59 \% $, respectively. While finding $ \lambda_{\mathrm{opt}} $ could be time-consuming, we notice that for a reasonable value of $ \lambda = \left\lbrace 0.05, 0.1 \right\rbrace $ the fully-digital and hybrid precoders behave remarkably compared to \texttt{\small{XHAUS}} while outperforming \texttt{\small{SING}} and \texttt{\small{RAND}}. 

\vspace{3mm}

\vspace{-3mm}
\section{Discussion and Future Work}
\label{sec:discussion}

\noindent{\textbf{Optimization Parameters:}}
Finding $ \lambda_{\mathrm{opt}} $ is challenging as it depends on the channel between the transmitter and receivers, as well as on the total interference. In order to circumvent this problem, a system could be trained using deep-learning to map the inputs to a suitable value of $ \lambda $. Other possible options are data clustering and multivariate regression. 

\noindent{\textbf{Hybrid Precoder Design:}}
Compared to most approaches on hybrid precoding (multi-user or multicast) in the literature, in this paper, the hybrid precoder is not obtained as an approximation of the optimal digital implementation. We design the precoder without any knowledge of the digital implementation. We presume that the performance of the hybrid precoder can be further boosted if a better initialization scheme is explored.

\noindent{\textbf{Hybrid Precoder Complexity:}}
The worst-case complexity of the hybrid precoder design, when using standard interior-point methods is $ N_{\mathrm{bis}_1} \mathcal{O} \left( G^3 N^6_\mathrm{tx} + K G N^2_\mathrm{tx} \right) + $ $ N_{\mathrm{bis}_2} \mathcal{O} \left( \left( G N^{\mathrm{RF}}_\mathrm{tx} \right)^6 + K \left( G N^{\mathrm{RF}}_\mathrm{tx} \right)^2 \right) + $ $ K N_{\mathrm{bis}_3} \mathcal{O} \left( N^6_\mathrm{rx} + N^2_\mathrm{rx} \right) $.

\noindent{\textbf{Optimal Scheduler:}}
Finding the optimal scheduler is intrinsically of combinatorial nature. Alternative relaxations of $ 0-1 $ parameters in \emph{Proposition 3} based on \texttt{log}, \texttt{exp}, and \texttt{arctan} functions could be further explored.


\vspace{-1mm}
\section{Conclusion}
\label{sec:conlusion}
We investigated the joint optimization of scheduling and multi-group multicast hybrid precoders to achieve ubiquitous low-latency mmWave communications in Industry 4.0 settings. We proposed a scheme based on alternate optimization, semidefinite relaxation and Cholesky matrix factorization to design the hybrid precoder. Also, we presented a novel scheduling formulation that takes into account the number of RF chains at the transmitter while minimizes the number of scheduling windows and channel correlation among the co-scheduled receivers. We corroborated through simulations that in terms of SINR the hybrid precoder can attain outstanding performance with a few number of RF chains. In terms of latency performance, the proposed \texttt{\small{HYDRAWAVE}} is capable of performing within $ 9.5 \% $ of the optimal \texttt{\small{XHAUS}} while exhibiting noticeable advantage over \texttt{\small{SING}} and \texttt{\small{RAND}}.

\vspace{-1mm}
\section{Acknowledgment} \label{acknowledgment}
The research is funded by the Deutsche Forschungsgemeinschaft (DFG) within the B5G-Cell project in SFB 1053 MAKI.

\begin{appendices}
\label{sec:appendix}

\setcounter{equation}{0}
\renewcommand{\theequation}{A.\arabic{equation}}
\section{Derivation of Proposition 1} \label{appA}
Let $ C^{\star}_k = C_{k \mid \mathcal{V}_{t'}, \mathbf{F}_{t'}, \mathbf{M}_{t'}, \mathbf{w}_k} $, $ \mathrm{SINR}^{\star}_k = \mathrm{SINR}_{k \mid \mathcal{V}_{t'}, \mathbf{F}_{t'}, \mathbf{M}_{t'}, \mathbf{w}_k} $ and $ \xi^{\star}_{t'} = \xi_{ t' \mid \mathcal{U}_{t'}, \mathbf{F}_{t'}, \mathbf{M}_{t'}, \left\lbrace \mathbf{w}_k \right\rbrace_{k \in \mathcal{U}_{t'}} } $. From (\ref{e4}), the transmission latency $ \xi^{\star}_{t'} $ with known $ \mathcal{U}_{t'}, \mathbf{F}_{t'}, \mathbf{M}_{t'}, \left\lbrace \mathbf{w}_k \right\rbrace_{k \in \mathcal{U}_{t'}} $ is defined as
\begin{subequations} \label{eA1}
	\begin{align}
		\xi^{\star}_{t'} = &
		\max_{k \in \mathcal{U}_{t'}} 
		\left\lbrace  
			\frac{B_k}{ \min \left\lbrace C^{\star}_k, C_\mathrm{max} \right\rbrace }
		\right\rbrace \label{eA1a}
		\\
		= &
		\max_{k \in \mathcal{U}_{t'}} 
		\left\lbrace 
			\max 
			\left\lbrace 
				\frac{B_k}{C^{\star}_k}, \frac{B_k}{C_{\mathrm{max}}} 
			\right\rbrace 
		\right\rbrace \label{eA1b}
		\\
		= &
		\frac{1}
		{
			\displaystyle \min_{k \in \mathcal{U}_{t'}} 
			\left\lbrace 
				\min 
				\left\lbrace 
					\frac{C^{\star}_k}{B_k}, \frac{C_{\mathrm{max}}} {B_k}
				\right\rbrace 
			\right\rbrace 
		} \label{eA1c}
		\\
		= &
		\frac{1}
		{
			\displaystyle \min_{k \in \mathcal{U}_{t'}} 
			\left\lbrace 
				\frac{\log_2 (1 + \beta \cdot \mathrm{SINR}^{\star}_k)}{B_k}, \frac{C_{\mathrm{max}}} {B_k}
			\right\rbrace 
		} \label{eA1d}
		\\
		= &
		\frac{1}
		{
			\log_2
			\left( 
			\displaystyle \min_{k \in \mathcal{U}_{t'}} 
			\left\lbrace 
				(1 + \beta \cdot \mathrm{SINR}^{\star}_k)^\frac{1}{B_k}, 2^{\frac{C_{\mathrm{max}}}{B_k}}
			\right\rbrace 
			\right) 
		}. \label{eA1e}
	\end{align}
\end{subequations}

Because (\ref{eA1c}) involves extremizations of the same type, they can be combined. When $ | \mathcal{V}_t | > 1 $, in general $ 1 + \beta \cdot \mathrm{SINR}^{\star}_k \leq 2^{C_{\mathrm{max}}} $, due to interference and restricted transmit/receive power. Further, since $ \log_2(\cdot) $ is monotonically increasing, then,
\begin{align} \label{eA2}
\scriptsize
	\xi^{\star}_{t'} \approxprop^{-1}
	\displaystyle \min_{k \in \mathcal{U}_{t'}} 
	\left\lbrace 
			(1 + \beta \cdot \mathrm{SINR}^{\star}_k)^\frac{1}{B_k}
	\right\rbrace .
\end{align}

\setcounter{equation}{0}
\renewcommand{\theequation}{B.\arabic{equation}}
\section{Derivation of Proposition 2} \label{appB}
Let $ \xi^{\star}_{t'} = \xi_{t' \mid \mathcal{U}_{t'}} $ and $ \mathrm{SINR}^{\star}_k = \mathrm{SINR}_{k \mid \mathcal{U}_{t'}} $ represent the latency during window $ t' $ and the SINR of receiver $ k \in \mathcal{U}_{t'} $, respectively when solely $ \mathcal{U}_t $ is known. Thus, $ \mathbf{F}_{t'}, \mathbf{M}_{t'}, \left\lbrace \mathbf{w}_k \right\rbrace_{k \in \mathcal{U}_{t'}} $ are be designed in order to minimize (\ref{eA1c}), i.e.,
\begin{align} \label{eB1}
	& \min_{\substack{
					\mathbf{F}_{t'} \in \Omega_F, 
					\mathbf{M}_{t'} \in \Omega_M \\ 
					\left\lbrace \mathbf{w}_k \right\rbrace_{k \in \mathcal{U}_{t'}} \in \Omega_W
	   }
	 }
	 \frac{1}
		 		{
			 		\displaystyle \min_{k \in \mathcal{U}_{t'}} 
			 		\left\lbrace  
				 		\min 
				 		\left\lbrace 
				 			\frac{C^{\star}_k}{B_k}, \frac{C_{\mathrm{max}}} {B_k}
				 		\right\rbrace 
			 		\right\rbrace, 
		 		} 	
\end{align} 
which is equivalent to 
\begin{subequations} \label{eB2}
	\begin{align} 		
		\overset{\mathrm{(B.1)}}{=} 
		& \max_{\substack{
						\mathbf{F}_{t'} \in \Omega_F, 
						\mathbf{M}_{t'} \in \Omega_M \\ 
						\left\lbrace \mathbf{w}_k \right\rbrace_{k \in \mathcal{U}_{t'}} \in \Omega_W
					   }
			 }
		 		\min_{k \in \mathcal{U}_{t'}} 
		 		\left\lbrace  
			 		\min 
			 		\left\lbrace 
			 			\frac{C^{\star}_k}{B_k}, \frac{C_{\mathrm{max}}} {B_k}
			 		\right\rbrace 
		 		\right\rbrace \label{eB2a}
		\\ 		
		\overset{\mathrm{(B.1)}}{=} 	 
		& \max_{\substack{
						\mathbf{F}_{t'} \in \Omega_F, 
						\mathbf{M}_{t'} \in \Omega_M \\ 
						\left\lbrace \mathbf{w}_k \right\rbrace_{k \in \mathcal{U}_{t'}} \in \Omega_W
					   }
			 }
		 		\min_{k \in \mathcal{U}_{t'}} 
		 		\left\lbrace  
			 		\frac{C^{\star}_k}{B_k}
		 		\right\rbrace \label{eB2b}
		\\ 		
		\overset{\mathrm{(B.1)}}{=} 	 
		& \max_{\substack{
						\mathbf{F}_{t'} \in \Omega_F, 
						\mathbf{M}_{t'} \in \Omega_M \\ 
						\left\lbrace \mathbf{w}_k \right\rbrace_{k \in \mathcal{U}_{t'}} \in \Omega_W
					   }
			 }
		 		\min_{k \in \mathcal{U}_{t'}} 
		 		\left\lbrace  
			 		\frac{\log_2 \left( 1 + \beta \cdot \mathrm{SINR}^{\star}_k \right) }{B_k}
		 		\right\rbrace \label{eB2c}
		\\ 		
		\overset{\mathrm{(B.1)}}{=} 	 
		& \max_{\substack{
						\mathbf{F}_{t'} \in \Omega_F, 
						\mathbf{M}_{t'} \in \Omega_M \\ 
						\left\lbrace \mathbf{w}_k \right\rbrace_{k \in \mathcal{U}_{t'}} \in \Omega_W
					   }
			 }
		 		\min_{k \in \mathcal{U}_{t'}} 
		 		\left\lbrace  
			 		\frac{\mathrm{SINR}^{\star}_k}{B_k}
		 		\right\rbrace \label{eB2d}
		\\
		\overset{\mathrm{(B.1)}}{=}
		& \max_{\substack{
						\mathbf{F}_{t'} \in \Omega_F, 
						\mathbf{M}_{t'} \in \Omega_M \\ 
						\left\lbrace \mathbf{w}_k \right\rbrace_{k \in \mathcal{U}_{t'}} \in \Omega_W
		   }
		 }
		 \min_{i \in \mathcal{V}_{t'}} \min_{k \in \mathcal{G}_i} \left\lbrace \frac{\mathrm{SINR}^{\star}_k}{B_k} \right\rbrace, \label{eB2e} 	 				 			 		
	\end{align}
\end{subequations}	
where (\ref{eB2a}) collapses to (\ref{eB2b}) because $ C_\mathrm{max} $ does not depend on $ \mathbf{F}_{t'}, \mathbf{M}_{t'}, \left\lbrace \mathbf{w}_k \right\rbrace_{k \in \mathcal{U}_{t'}} $. In (\ref{eB2c}), since $ \log_2 \left( 1 + \beta \cdot \mathrm{SINR}^{\star}_k \right) $ is an injective mapping of $ \mathrm{SINR}^{\star}_k $, it is equivalent to (\ref{eB2d}). Further, (\ref{eB2e}) is an alternative notation for (\ref{eB2d}).

\end{appendices}

\bibliographystyle{IEEEtran}
\vspace{-2mm}
\bibliography{ref}
\end{document}